\documentclass{aa} 

\usepackage{graphicx}
\usepackage{txfonts}
\usepackage{xcolor}

\newcommand{\cxo}{{\it Chandra}}

\newcommand{\xmm}{{\it XMM-Newton}}
\newcommand{\xmms}{{\it XMM}}
\newcommand{\icmz}{{\texttt{i(cm)z}}}

\begin{document}

   \title{\icmz, a semi-analytic model for the thermodynamic properties
   in galaxy clusters: Calibrations with mass and redshift, \\
   and implication for the hydrostatic bias}
\titlerunning{Calibration of \icmz}

  \author{S. Ettori\inst{1,2}
          \and
          L. Lovisari\inst{1,3} 
          \and
          D. Eckert\inst{4}  
        }

   \institute{INAF, Osservatorio di Astrofisica e Scienza dello Spazio, via Piero Gobetti 93/3, 40129 Bologna, Italy
        \and
        INFN, Sezione di Bologna, viale Berti Pichat 6/2, 40127 Bologna, Italy
        \and
        Center for Astrophysics $|$ Harvard $\&$ Smithsonian, 60 Garden Street, Cambridge, MA 02138, USA
        \and 
         Department of Astronomy, University of Geneva, ch. d’Ecogia 16, 1290 Versoix, Switzerland
        }
        \authorrunning{Ettori et al.}

  \abstract
   {In the self-similar scenario for galaxy cluster formation and evolution, the thermodynamic properties of the X-ray emitting plasma can be predicted in their dependencies on the halo mass and redshift only. However, several departures from this simple self-similar scenario have been observed. }
   {We show how our semi-analytic model \icmz, which modifies the self-similar predictions through two temperature-dependent quantities, the gas mass fraction $f_g = f_0 T^{f_1} E_z^{f_z}$ and the temperature variation $f_T = t_0 T^{t_1} E_z^{t_z}$, can be calibrated to incorporate the mass and redshift dependencies. }
   {We used a published set of 17 scaling relations to constrain the parameters of the model. We were subsequently able to make predictions as to the slope of any observed scaling relation within a few percent of the central value and about one $\sigma$ of the nominal error.
   Contextually, the evolution of these scaling laws was also determined, with predictions within $1.5 \sigma$ and within 10 percent of the observational constraints. 
   Relying on this calibration, we have also evaluated the consistency of the predictions on the radial profiles with some observational datasets. 
   For a sample of high-quality data (X-COP), we were able to constrain a further parameter of the model, the hydrostatic bias $b$.}
   {By calibrating the model versus a large set of X-ray scaling laws, we have determined that (i) the slopes of the temperature dependence are $f_1 = 0.403 \; (\pm 0.009)$ and $t_1 = 0.144 \; (\pm 0.017)$; and that (ii) the dependence upon $E_z$ are constrained to be $f_z = -0.004 \; (\pm 0.023)$ and $t_z = 0.349 \; (\pm 0.059)$.
   These values, which are inserted in the scaling laws that propagate the mass and redshift dependence to the integrated quantities, permit one to estimate directly how the normalizations of a given quantity $Q_{\Delta}$ changes as a function of the mass (or temperature) and redshift halo in the form $Q_{\Delta} \sim M^{a_M} \, E_z^{a_z} \sim T^{a_T} \, E_z^{a_{Tz}}$, which is in very good agreement with the current observational constraints.
   When applied to the best spatially resolved data, we obtained estimates of the hydrostatic bias $b$ that are lower than, but still comparable with, the results obtained by other, more standard, means. We conclude that the calibrated semi-analytic model \icmz\ is able to make valuable predictions on the slope and redshift evolution of the X-ray scaling laws, and on the expected radial behavior of the thermodynamic quantities, including any possible hydrostatic bias. }
   {}

   \keywords{galaxy: clusters: general --
             galaxy: clusters: intracluster medium -- 
             X-rays: galaxy: clusters
               }

   \maketitle
%
\section{Introduction}

Galaxy groups and clusters are overdensities in the cosmic field where gravity, mostly produced from a still-unknown dark component, rules and regulates most of their appearance. The baryons flow across these halos, and the largest part of those is heated up during this circulation process reaching temperatures that make them shine in X-ray and affect the distribution of the cosmic microwave background (CMB) photons at millimeter wavelengths.
Being gravitationally bound and dark matter dominated, these massive halos have observed properties that depend, in general and at first approximation, on the mass and redshift of the halo only \citep[see e.g.,][]{voit05,allen11}.

However, several departures from this simple, self-similar scenario have been observed and are thought to be consequences of the time and mass-dependent feedback from active galactic nuclei (AGNs) and star formation, with a greater impact on systems where this feedback contributes in a non-negligible way to the overall energy budget. This seems to occur more regularly and systematically in halos at lower masses \citep[$\sim 10^{13} - 10^{14} M_{\odot}$; see e.g.,][]{eck21,lov21,opp21}, where this extra energy becomes comparable to the total binding one. 
The action of this feedback over the cosmic life of these structures impacts  the overall baryon distribution in such a way that the ideal condition of an isolated ``closed box,'' where the total baryon mass fraction matches the cosmological value, is reached well beyond any nominal virial radius \citep[at $>6 R_{500}$; see e.g.,][]{angelinelli22}.

In the following analysis, we refer to masses $M_{\Delta}$ that are associated with a given overdensity $\Delta$ as 
$M_{\Delta} = 4/3 \, \pi \, \Delta \, \rho_{\rm c,z} R_{\Delta}^3$, 
where $\rho_{\rm c,z} = 3 H_z^2 / (8 \pi G)$ is the critical density of the universe at the observed redshift $z$ of the cluster,
$G$ is the universal gravitational constant, and $H_z = H_0 \, \left[\Omega_{\Lambda} +\Omega_{\rm m}(1+z)^3\right]^{0.5}
= H_0 \, E_z$ is the value of the Hubble constant at the same redshift.
For the $\Lambda$ Cold Dark Matter (CDM) model, we adopt the cosmological parameters
$H_0=70$ km s$^{-1}$ Mpc$^{-1}$ and $\Omega_{\rm m} = 1 - \Omega_{\Lambda}=0.3$.

\section{The \icmz\ model}
\label{sect:icmz}

The semi-analytic model \icmz\ has been presented in \cite{ettori20} (hereafter E20).
It generates radial profiles of the thermodynamic quantities of galaxy groups and clusters by putting a universal pressure profile in hydrostatic equilibrium with a gravitational potential described via a concentration-mass relation.
In detail, the procedure combines the following steps:
   \begin{itemize}
   \item as input, a total mass ($M_{500}$ in our analysis) and a redshift $z$ are provided;

   \item a concentration-mass-redshift relation is assumed, 
   \begin{equation}
   \log c_{200} = A +B \log (M_{200} / 10^{12} M_{\odot} h_{100}^{-1}),
   \label{eq:cmz}
   \end{equation} 
   with $B = -0.101 +0.026 \, z$ and $A = 0.520 +(0.905-0.520) \exp(-0.617 z^{1.21})$ \citep{dutton14}; alternative forms can be considered \citep[e.g.,][]{bha13,lud16}; the conversion from $M_{200}$ to $M_{500}$ is obtained through iteration assuming a Navarro, Frenk \& White \citep{nfw97} profile;
 
   \item a functional form is adopted for the universal electronic pressure profile,
   \begin{equation} P_{\rm e} = P_{500} P_r,  \label{eq:puniv} \end{equation} 
   with $P_{500} = 1.65 \times 10^{-3} \left(M_{500} / 3 \times 10^{14} h_{70}^{-1} M_{\odot} \right)^{2/3 +\alpha_M}$ keV cm$^{-3} \, E_z^{8/3} \, h_{70}^2$ and 
   $P_r =  P_0 \, (c_{500} x)^{-\gamma} \,  \left[1 + (c_{500} x)^{\alpha} \right]^{(\beta- \gamma) / \alpha}$ with $x=r / R_{500}$, the parameters $(P_0, c_{500}, \gamma, \alpha, \beta)$ equal to $(8.403, 1.177, 0.308, 1.051, 5.491)$, and $\alpha_M = 0.12$, accounting for the observed deviation from the standard self-similar scaling, set as in \cite{arnaud10} \citep[alternative values for the same model or different models are available in, e.g.,][]{planck13,ghi19univ,ghi19poly,sayers22};
 
   \item the hydrostatic equilibrium equation is considered and inverted to recover the electron density
   \begin{equation} 
   n_{\rm e} = -\frac{dP_{\rm e}}{dr} \frac{r^2}{\mu \, m_a \, G \, M_{\rm HE}},
   \label{eq:mhe}
   \end{equation}
   where $\mu \approx 0.6$ is the mean molecular weight of the gas, $m_a$ is the atomic mass unit of $1.66 \times 10^{-24}$ g, and $G$ is the universal gravitational constant;
   
   \item the gas temperature profile is obtained using the perfect gas law and dividing the electron pressure in eq.~\ref{eq:puniv} and the electron density in eq.~\ref{eq:mhe}
   \begin{equation} 
   T_{\rm gas, 3D} = P_{\rm e} / n_{\rm e};
   \label{eq:tgas}
   \end{equation}
   
   \item in the case a hydrostatic bias $b = 1-M_{\rm HE}/M$ is considered (see Sect.~\ref{sect:bhe}), the input mass is corrected by a factor $1/(1-b)$ only for eq.~\ref{eq:cmz}, leaving the nominal, uncorrected value $M_{\rm HE}$ in all the remaining cases (e.g., for the determination of $R_{500}$ or for the normalization of $P_{500}$) to mimic what observers would do.

   \end{itemize}
From these radial profiles, the integrated quantities are recovered numerically over the volume of interest (see E20 for further details).

\begin{figure*}[ht]
\includegraphics[trim=10 0 0 220,clip,width=\hsize]{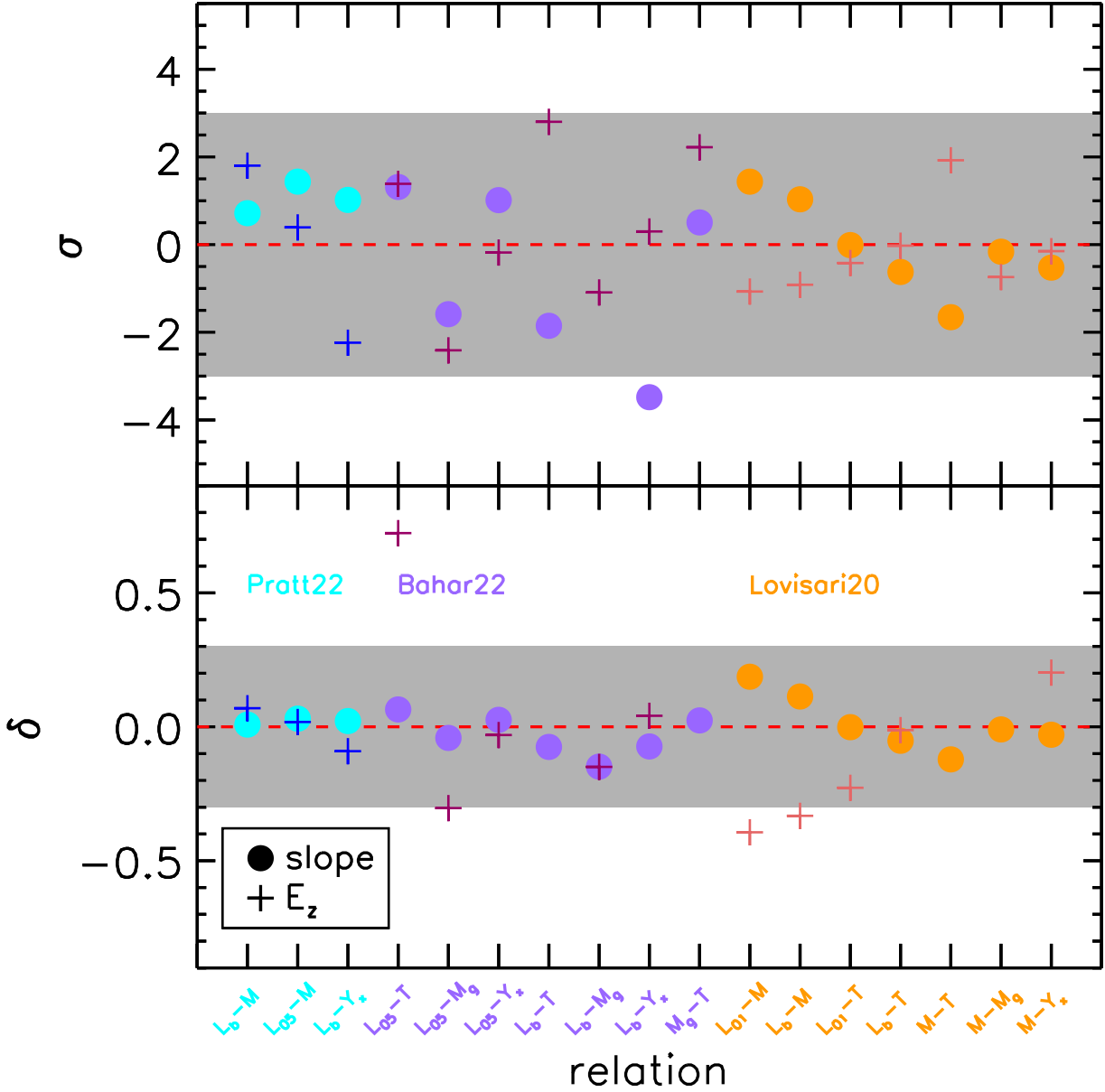} 
\caption{Differences between the predicted and observed values of the slope (filled circle) and the redshift dependence (cross) of the listed scaling relations.
(Top panel) Deviations in terms of $\sigma = (O-M) / \epsilon_O$, where $O$ is the estimate constrained from the observations with error $\epsilon_O$ and $M$ is the predicted value; (bottom panel) systematic deviation $\delta = (O-M) / M$. 
Shaded regions represent $[-3, +3] \, \sigma$ and $[-0.3, +0.3] \, \delta$ in the top and bottom panel, respectively.
Labels indicate the following: luminosity, either bolometric ($L_{\rm b}$) or in bands (0.5-2 keV, $L_{05}$; 0.1-2.4 keV, $L_{01}$); total ($M$) and gas ($M_{\rm g}$) mass; and temperature $T$ and X-ray $Y_x = M_{\rm g} \, T$.  
} \label{fig:scalaw}
\end{figure*}

\subsection{Departures from self-similarity: The temperature (mass) and redshift dependence}
\label{sect:SSmod}

In gravity-dominated systems, the so-called self-similar relations between the integrated quantities, such as luminosity, gas mass, and temperature, hold. However, departures from this self-similar scenario are expected, and they are observed now systematically, going down in mass, where radiative and feedback processes impact the intra-cluster medium (ICM) distribution and the action of the gravity alone, inducing deviations in the gas distribution both as a function of radius and in the single power-law behavior between integrated quantities.

As described in our previous work \citep[see e.g.,][]{ettori15,ettori20}, we account for these departures by introducing temperature- (mass-)dependent quantities such as the gas mass fraction $f_g$, and the variation between the global spectroscopic temperature $T_{\rm spec}$ and its value at $R_{500}$, $f_T = T(R_{500}) / T_{500} \times T_{500} / T_{\rm spec} = T(R_{500}) / T_{\rm spec}$. These quantities allow us to express the corrections needed to accommodate for the observed deviations from the self-similar scenario, both radially and in the scaling relations in the form of a power-law dependence on either the gas temperature or the halo mass. 

Following E20, we use the following notation hereafter:
\[
      \begin{array}{lp{0.8\linewidth}}
         M           & total mass used in eq.~\ref{eq:cmz}   \\
         M_{\rm HE}  & hydrostatic mass used in eq.~\ref{eq:puniv}, \ref{eq:mhe}  \\
         b = 1-M_{\rm HE}/M  & hydrostatic mass bias  \\
         f_T = T(R_{500}) / T_{\rm spec} = t_0 T^{t_1} &  variation in temperature \\
         f_g = M_g / M_{\rm HE} = f_0 T^{f_1} & gas mass fraction dependence on $T$. \\
      \end{array}
\]
\noindent This will allow us to write a modified, more flexible, version of the standard self-similar scaling laws.
We refer readers to E20 for a further detailed treatment of the normalizations $t_0$ and $f_0$.
For the sake of clarity, as we also discuss in E20, we note that the X-ray-measured gas mass fraction $f_g = M_g / M_{\rm HE}$ is related to the ``true'' gas mass fraction (i.e., the one unbiased from the hydrostatic bias and the clumping factor) by the relation 
$f_{\rm gas, true} = M_g / M_{tot} = (1-b)^{2/3} \, f_g = (1-b)^{2/3} \,  C^{0.5} \, f_{nc}$, where we assume that $M_g$ is proportional to $R_{\Delta}$ and $R_{\Delta, \rm HE} = (1-b)^{1/3} R_{\Delta, tot}$,
$f_{nc}$ is the clumping-free gas mass fraction, and $C = <n_{\rm gas}^2> / <n_{\rm gas}>^2$ is the clumping factor. 

In E20 \citep[see also][]{ettori15}, we developed all the equations in function of the gas temperature and quote the rescaling also as a function of the total mass only for the normalizations of the thermodynamic quantities \citep[see Table~4 in][]{ettori20}.
For the sake of completeness and because it is useful for the new calibrations that we discuss in the present work, we have rewritten those relations, focusing only on their dependency on $T$ (or $M$):
\begin{align}
\mathcal{M} & \sim (1-b)^{-1} \, f_T ^{3/2} \, \mathcal{T}^{3/2} \sim \mathcal{T}^{3/2 +3/2 t_1} \nonumber \\
\mathcal{M}_g & \sim f_g \, f_T ^{3/2} \, \mathcal{T}^{3/2} \sim \mathcal{T}^{3/2 +3/2 t_1 +f_1} \nonumber \\
\mathcal{L} & \sim f_g^2 \, f_T ^{3/2} \, \mathcal{T}^2 \sim \mathcal{T}^{3/2 +c + 3/2 t_1 +2 f_1} \nonumber \\
\mathcal{T} & \sim \mathcal{M}^{2/3 \times 1/(1+t_1)}.
\label{eq:scalaw}
\end{align}
Here, we have defined the following quantities: the total mass $\mathcal{M} \equiv E_z M_{tot} / M_0$ ($M_0 = 5 \times 10^{14} M_{\odot}$), the gas mass $\mathcal{M}_g \equiv E_z M_g / M_{g,0}$ ($M_{g,0} = 5 \times 10^{13} M_{\odot}$), a bolometric luminosity $\mathcal{L} \equiv E_z^{-1} L / L_0$ ($L_0 = 5\times10^{44}$ erg s$^{-1}$), and a gas temperature $\mathcal{T} \equiv k_B T_{\rm spec} / T_0$ ($T_0 =$ 5 keV). 
The exponent $c$ indicates the temperature dependence of the cooling function and is equal to $0.44$ (for the pseudo-bolometric 0.01-100 keV band), $-0.11$ (0.1--2.4 keV band), and $-0.09$ (0.5--2 keV band), with a slight dependence on the metallicity and on the abundance table for the reference assumed \citep[0.3 times the solar value in Asplund et al. 2009 in our case; see also][and some further considerations in Sect.~\ref{sect:limits}]{ettori15,lov21}. 

\subsection{Evolution with redshift}
\label{sect:zevol}

So far, the redshift evolution has been considered only in the part related to the self-similar model through the quantity $E_z$.
We have expand this approach by also allowing a redshift evolution to the $f_g$ and $f_T$ parameters in the following form: 
$f_g = f_0 T^{f_1} E_z^{f_z}$ and $f_T = t_0 T^{t_1} E_z^{t_z}$, with two extra parameters $f_z$ and $t_z$.

By propagating the dependence through the scaling laws, we obtain 
\begin{align}
M & \sim \mathcal{T}^{3/2 +3/2 t_1} \; E_z^{3/2 t_z-1} \nonumber \\
M_g & \sim \mathcal{T}^{3/2 +3/2 t_1 +f_1} \; E_z^{f_z-3/2t_z-1} \nonumber \\
L & \sim \mathcal{T}^{3/2 +c + 3/2 t_1 +2 f_1} \; E_z^{2f_z +3/2t_z +1} \nonumber \\
T & \sim M^{2/3 \times 1/(1+t_1)} \; E_z^{(3/2-t_z)/(1+t_1)},
\label{eq:scalaw_ez}
\end{align}
where we indicate, on the right side, the entire contribution assigned to the evolution in the form of $E_z$; that is to say, the quantities on the left side are the ones actually measured and are the same as the ones in eq.~\ref{eq:scalaw} once the dependence on $E_z$ is moved to the right side of the equation (e.g., $M =\mathcal{M} E_z^{-1}$, $L = \mathcal{L} E_z$).

\section{Calibration of the model parameters with observed scaling laws}
\label{sect:calib}

The \icmz\ model assumes some power-law relations among the integrated quantities. These relations are presented in eq.~\ref{eq:scalaw}, and their redshift evolution in eq.~\ref{eq:scalaw_ez}. Any new relation can be obtained by propagation of those, allowing us to predict the expected values of the slope (and redshift evolution) for any scaling relation as a function of only two parameters, $f_1$ and $t_1$, for the slope, and two others for the $z    $ evolution, $f_z$ and $t_z$.

To calibrate these values, we have collected from recent literature the best-fit results for the slopes and redshift evolution of several scaling relations.
To make full use of the entire set of considered relations and of the statistical uncertainties on their best-fit constraints, we proceeded as follows:
\begin{enumerate}[1.]
    \item we randomly drew, from a flat distribution between 0 and 1, two numbers representing the exploring set $i$ of values $\{f_{1i},t_{1i}\}$;
    \item for any scaling relation $j$ considered, we assigned a probability $P_{ij} = \exp{(-(O_j-M_j)^2/\epsilon_{O_j}^2}$ to each set $i$ of values, where $O_j$ is the observational constraint on the slope with an associated error $\epsilon_{O_j}$ and $M_j$ is the predicted value provided from the set $\{f_{1i},t_{1i}\}$;
    \item for each set $i$ of values, we built a cumulative probability $P_i = \prod_j P_{ij}$, where the product was done over the entire collection of scaling relations $j$ under consideration that are considered independent of each other here;
    \item we repeated the procedure $N$ (with $N\sim 10^6$) times and estimate the mean and the dispersion of the values as $\bar{f_1} = \sum_i w_i f_{1i}$ and $\sigma_{f_1} = \left( \sum w_i (f_{1i}-\bar{f_i})^2 \right)^{1/2}$, with the weights $w_i = P_i / \sum P_i$. The same calculation was performed to estimate $\bar{t_1}$ and $\sigma_{t_1}$.
\end{enumerate}

We selected any scaling law where the normalization, slope, and redshift evolution were considered as free parameters. 
If relations for more subsamples are available, we would use the one with the largest coverage in mass and redshift.
We used a total number of 17 scaling laws extracted from three different samples:
\begin{description}
\item[{\bf \cite{lovisari20}}] discuss the integrated properties of 120 galaxy clusters in the Planck Early Sunyaev-Zel'dovich (ESZ) sample spanning the mass range $2.4-17.6 \times 10^{14} M_{\odot}$ in the redshift interval $0.059 < z < 0.546$; 
\item[{\bf \cite{bahar22}}] present the analysis of scaling relations for a sample of 265 clusters extracted from the {\it eROSITA} Final Equatorial-Depth Survey (eFEDS) field with a contamination level of $<10$\% (including AGNs and spurious fluctuations) and covering ranges in total mass of $6.9 \times 10^{12} M_{\odot} < M_{500} < 7.8 \times 10^{14} M_{\odot}$ (with 68 systems at masses below $10^{14} M_{\odot}$ representing one of the largest group samples detected uniformly to date) and in redshift of $0.02 < z < 0.94$; 
\item[{\bf \cite{pratt22}}] studied the scaling laws of 93 SZ-selected objects with \xmm\ exposures spanning a mass range of $M_{500} = 0.5 - 20 \times 10^{14} M_{\odot}$ and lying at redshifts $0.05 < z < 1.13$.
\end{description}

We noticed that no specific information on the dynamical state of the objects of these samples is available. 
However, given the SZ-selected samples of \cite{lovisari20} and \cite{pratt22}, we expect a contribution from cool-core (more relaxed) objects to be lower than in samples selected through their X-ray emission \citep[in particular, when the core contribution is not compensated for; see e.g.,][]{eckert11}, and to be around $\sim$ 30-50\% of the total \citep[see e.g.,][]{rossetti16,nurgaliev17,andrade17}.
For what concerns the sample in \cite{bahar22}, the investigation of the morphological properties in a more extended eFEDS sample of 325 objects presented in \cite{ghi22} indicates that about 30-40\% of the clusters are dynamically relaxed.
For the sake of simplicity, but not affecting the overall efficiency in reproducing the observed datasets, we have not considered any dependency on the dynamical state in the current parametrization of the \icmz\ model (see more comments in Sect.~ \ref{sect:limits}).

The final outcomes of this process are the following constraints on $f_1$ and $t_1$:  
\begin{eqnarray}
    \bar{f_1} & = & 0.403 \; (\sigma_{f_1} =  0.009) \nonumber \\
    \bar{t_1} & = & 0.144 \; (\sigma_{t_1} = 0.017),
\label{eq:f1t1}
\end{eqnarray}
with an associated mild correlation coefficient $\rho = {\rm cov}(f_1, t_1) / (\sigma_{f_1}\sigma_{t_1})$ of 0.44.

When the analysis was repeated for each of the three samples independently, we measured the following: 
$f_1 = 0.45 \; (0.04)$ and $t_1 = 0.05 \; (0.03)$, only using the seven relations from \cite{lovisari20};
$f_1 = 0.47 \; (0.03)$ and $t_1 = 0.25 \; (0.07)$, only using the three relations from \cite{pratt22}; and
$f_1 = 0.22 \; (0.02)$ and $t_1 = 0.40 \; (0.04)$, with the seven scaling laws from \cite{bahar22}.
The best-fit results presented in equation~\ref{eq:f1t1} represent a sort of weighted mean of these values.
We noticed that SZ-selected samples (such as the ones in Lovisari et al. 2020 and Pratt et al. 2022) seem to prefer $f_1 > t_1$, whereas the reverse can be observed in the X-ray-selected sample of \cite{bahar22}.
With our goal being to provide a first, exhaustive calibration of the physically meaningful parameters of the \icmz\ model $f_g$ and $f_T$, we have made use of the entire collection of scaling relations, assigning them equal weight (and with a statistical weight defined by their corresponding error) once combined to obtain the best-fit results presented in equation~\ref{eq:f1t1}.

\begin{figure}[ht]
\includegraphics[trim=30 40 50 230,clip,width=\hsize]{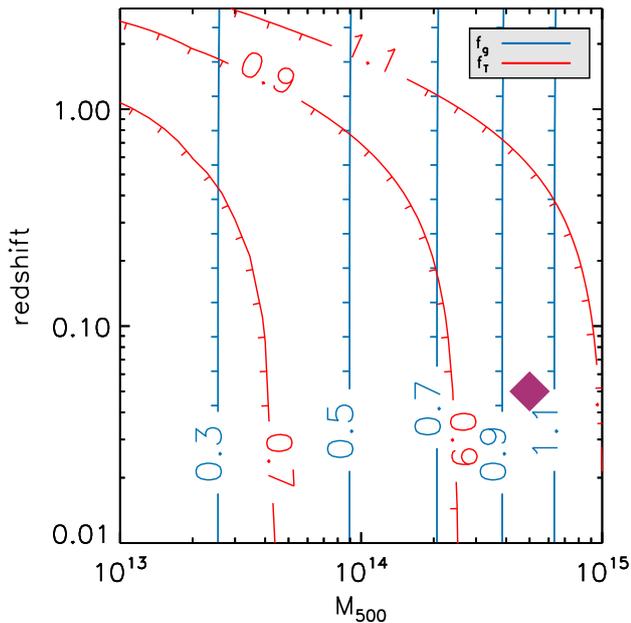} 
\caption{Predicted dependence on mass and redshift of the quantities $f_g$ (blue contours) and $f_T$ (red contours) renormalized to 1 at $(M_{500}, z) = (5 \times 10^{14} M_{\odot}, 0.05)$ (purple diamond). 
} \label{fig:ftfg}
\end{figure}

This procedure has also been validated by comparing the derived and the observed constraints of the slopes of the scaling laws considered. In Fig.~\ref{fig:scalaw}, we plot the deviations between the observed constraints and the best-fit results of eq.~\ref{eq:f1t1}. We show both the statistical deviations, as the difference between the central values in terms of the quoted error, $\sigma = (O-M)/\epsilon_O$, and the systematic one as the difference between the central values divided by the model prediction, $\delta = (O-M)/M$.
We measured a median (and first and third quartiles) of $\sigma = -0.01 \, (-1.59, 1.02)$  and of $\delta = 0.0 \, (-0.05, 0.03)$ on the measurements of the slope.

Using the observed limits on the evolution of the scaling relations, together with the constraints obtained above on $f_1$ and $t_1$, we put constraints on the redshift dependency of $f_g$ and $f_T$ (see Sect.~\ref{sect:zevol}):
\begin{eqnarray}
    \bar{f_z} & = & -0.004 \; (\sigma_{f_z} =  0.023) \nonumber \\
    \bar{t_z} & = & 0.349 \; (\sigma_{t_z} = 0.059),
\label{eq:fztz}
\end{eqnarray}
with a correlation coefficient of 0.63.
Adding this contribution to the evolution improved the modeling of the observed constraints.
We measured $\sigma = 0.16 \, (-1.79, 0.67)$ and $\delta = -0.08 \, (-0.43, 0.17)$ when these extra parameters were not included and the entire evolution is described by the modified self-similar model.
When we also propagated $f_z$ and $t_z$, we obtained
$\sigma = -0.15 \, (-0.92, 1.39)$ and $\delta = -0.03 \, (-0.30, 0.07)$.

The best-fit results presented in equations~\ref{eq:f1t1} and \ref{eq:fztz} provide a full description of the behavior of the physically meaningful parameters $f_g$ and $f_T$ of the \icmz\ model. 
As we show in Fig.~\ref{fig:ftfg}, we can recover their expected dependencies on the halo mass and redshift, suggesting that the gas fraction $f_g$ does not change with cosmic time, whereas a mild positive evolution is predicted for $f_T$, and that $f_g$ decreases significantly with $M_{500}$ (by 50\% going down in mass by an order of magnitude), and more severely than $f_T$.

\begin{table*}[ht]
\centering 
\caption{Dependences of the characteristic physical scales on the temperature, mass, and redshift (as $E_z^{\alpha_z}$) in the self-similar and  \icmz\ models.}
\begin{tabular}{c|cc|ccc}
\hline
Quantity  &  $f(M, z)$  &  $f(f_{\rm gas})$ & $f(T, z)$  & $f(M, z)$  \\
          &  \multicolumn{2}{c|}{self-similar} & \multicolumn{2}{c}{\icmz} \\
\hline 
\rule{0pt}{2.5ex} $T_{\Delta}$ & $M^{2/3} \, E_z^{2/3}$ & $f_{\rm gas}^0$ & $T^{1+t_1} \, E_z^{t_z}$ &  
$M^{2/3} \, E_z^{2/3}$  \\
\rule{0pt}{2.5ex} $n_{\Delta}$ & $E_z^2$ & $f_{\rm gas}^1$ & $T^{f_1} \, E_z^{2+f_z}$ & 
$M^{2/3 \, f_1/(1+t_1)} \, E_z^{2+f_z+f_1 (2/3-t_z)/(1+t_1)}$  \\
\rule{0pt}{2.5ex} $P_{\Delta}$ & $M^{2/3} \, E_z^{8/3}$  & $f_{\rm gas}^1$ & $T^{1 + t_1 + f_1} \, E_z^{2+f_z+t_z}$ & 
$M^{2/3 \, +2/3 f_1/(1+t_1)}  \, E_z^{8/3+f_z +f_1 (2/3-t_z)/(1+t_1)}$   \\
\rule{0pt}{2.5ex} $K_{\Delta}$ & $M^{2/3} \, E_z^{-2/3}$  &  $f_{\rm gas}^{-2/3}$ & $T^{1 + t_1 -2/3 f_1} \, E_z^{-4/3 -2/3 f_z +t_z}$ & 
$M^{2/3 \, -4/9 f_1/(1+t_1)} \, E_z^{-2/3 -2/3 f_z -2/3 f_1 (2/3-t_z)/(1+t_1)}$  \\
\hline
\end{tabular}
\tablefoot{The basic equations are $T_{\Delta} = f_T T \sim T^{1+t_1} \, E_z^{t_z} = (E_z M)^{2/3} = (E_z^3 R^3)^{2/3}$ and $n_{\Delta} \sim \Delta \rho_{cz} \sim f_g E_z^2 \sim T^{f_1} E_z^{2+f_z}$. All the other relations were obtained by combinations of those. 
}\label{tab:normz}
\end{table*}

\begin{table*}[ht]
\caption{Best-fit results for $Q_{\Delta} \sim \, M^{a_M} \, E_z^{a_z} \sim \, T^{a_T} \, E_z^{a_{Tz}}$.}
\begin{center} \begin{tabular}{ccccc } \hline
 $Q_{\Delta}$ & $a_M$ & $a_z$ & $a_T$ & $a_{T, z}$ \\ \hline
 $T_{\Delta}$ & $2/3$ $[2/3]$ & $2/3$ $[2/3]$  & $1.14 \; (0.02)$ $[1]$ & $0.35 \; (0.06)$ $[0]$ \\
 $n_{\Delta}$ & $0.23 \; (0.01)$ $[0]$ & $2.11 \; (0.03)$ $[2]$ & $0.40 \; (0.01)$ $[0]$ & $2.00 \; (0.02)$ $[2]$ \\
 $P_{\Delta}$ & $0.90 \; (0.01)$ $[2/3]$ & $2.78 \; (0.03)$ $[8/3]$ & $1.55 \; (0.02)$ $[1]$ & $2.35 \; (0.06)$ $[2]$ \\
 $K_{\Delta}$ & $0.51 \; (0.01)$ $[2/3]$ & $-0.74 \; (0.02)$ $[-2/3]$ & $0.88 \; (0.02)$ $[1]$ & $-0.98 \; (0.06)$ $[-4/3]$ \\
\hline \end{tabular} \end{center}
\tablefoot{Mean and dispersion for the parameters $a_M$, $a_z$, $a_T$, and $a_{Tz}$ obtained from the constraints on $f_g \sim T^{f_1} E_z^{f_z}$ and $f_T \sim T^{t_1} E_z^{t_z}$ in equations~\ref{eq:f1t1} and \ref{eq:fztz}. 
In the square brackets, we quote the self-similar predictions.
We note that, in combination with the relations in Tab.~\ref{tab:normz}, other rescaling can be obtained. For example, the emission measure $EM \approx \int n^2 dl$, which can be directly reconstructed from the observed X-ray surface brightness, is expected to scale as 
$M^{1/3 +4/3 f_1/(1+t_1)} E_z^{10/3 +2 f_z +2 f_1 (2/3-t_z)/(1+t_1)} \sim M^{0.80 (0.01) \, [1/3]} E_z^{3.55 (0.06) \, [10/3]}$ and as $T^{1/2 +2 f_1 +t_1/2} E_z^{3 +2 f_z +t_z/2} \sim T^{1.38 (0.02) \, [1/2]} E_z^{3.17 (0.05) \, [3]}$.
} \label{tab:normz_bestfit}
\end{table*}

\begin{figure*}[ht]
\centering
\includegraphics[trim=0 45 0 560,clip,width=\hsize]{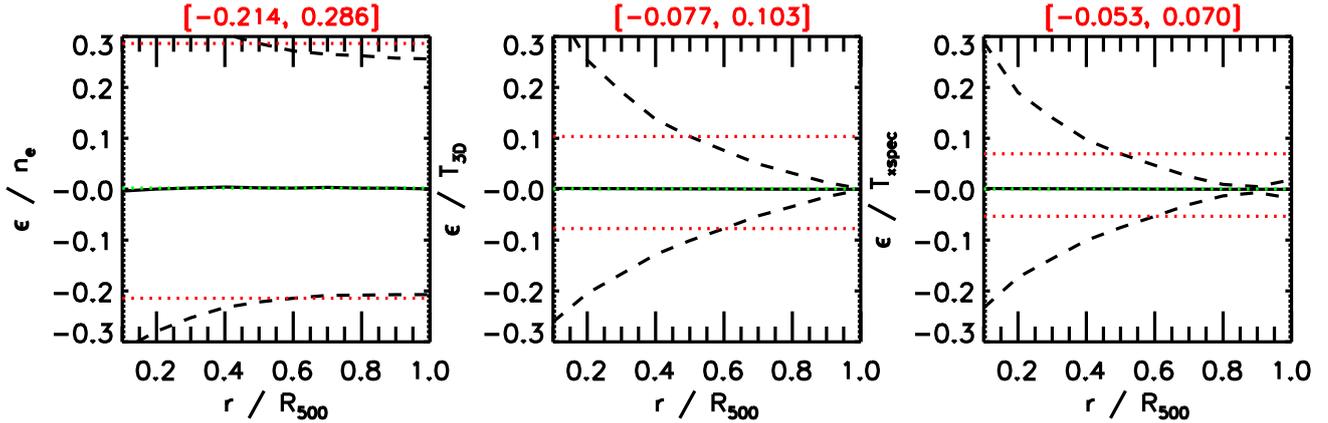} 
\caption{Radial profiles of the relative uncertainties ($\epsilon/Y-1$) on (left) gas density, (center) 3D gas temperature, and (right) projected spectroscopic-like temperature due to (i) the errors on the assumed parameters that describe the mass dependence, (ii) the scatter in the $c-M-z$ relation, and (iii) the intrinsic scatter on the ``universal'' pressure profile. 
Red (green) dotted lines indicate the median values of the 16th and 84th percentiles (and of the medians) estimated at each radius.  
The case refers to the input values $(M_{500}, z)=(4.6 \times 10^{14} M_{\odot}, 0.6)$, which represents the median values of the S18 sample.
} \label{fig:syst}
\end{figure*}

\begin{figure*}[ht]
\includegraphics[trim=10 10 70 220,clip,width=0.5\hsize]{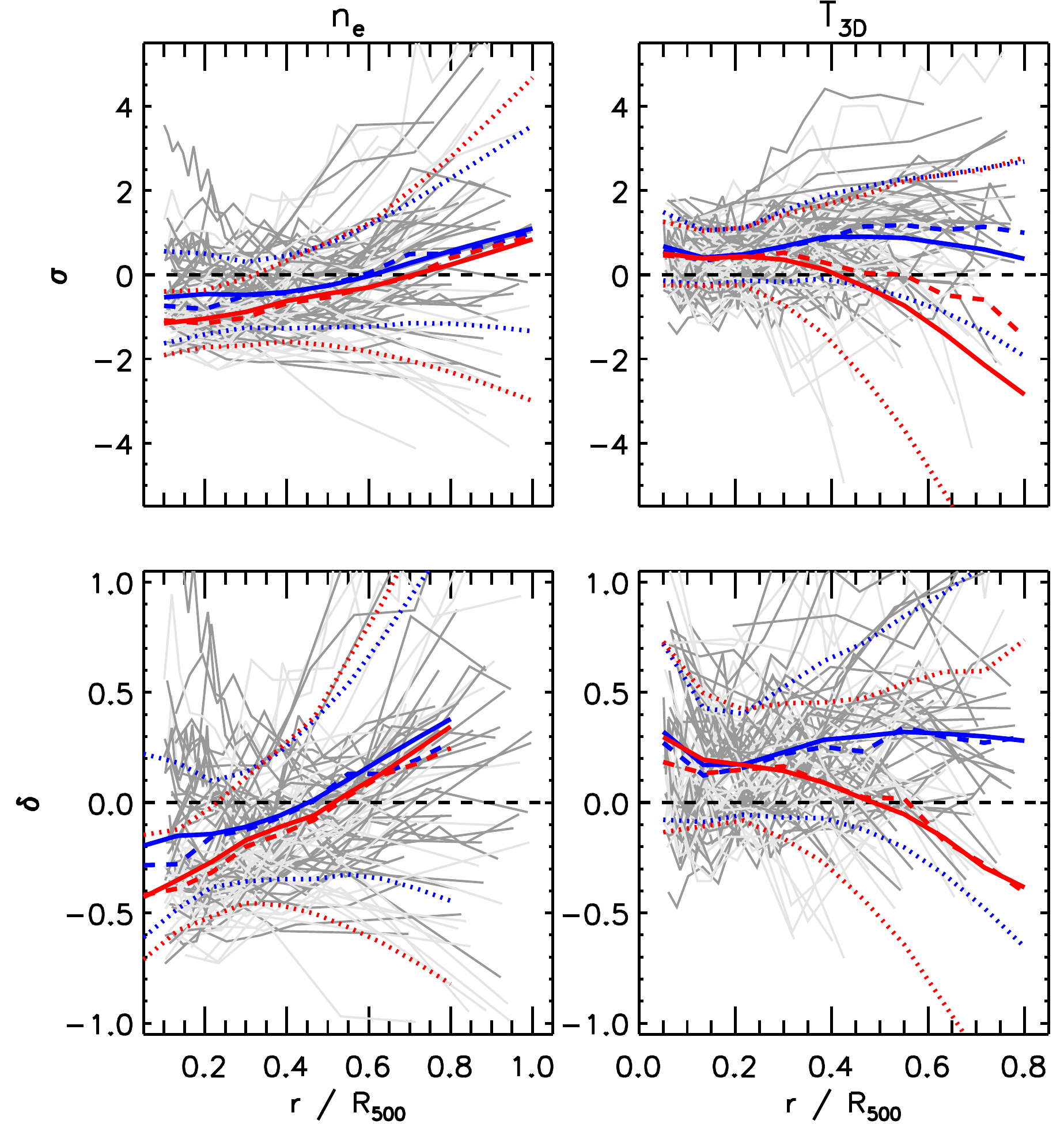}
\includegraphics[trim=10 10 70 220,clip,width=0.5\hsize]{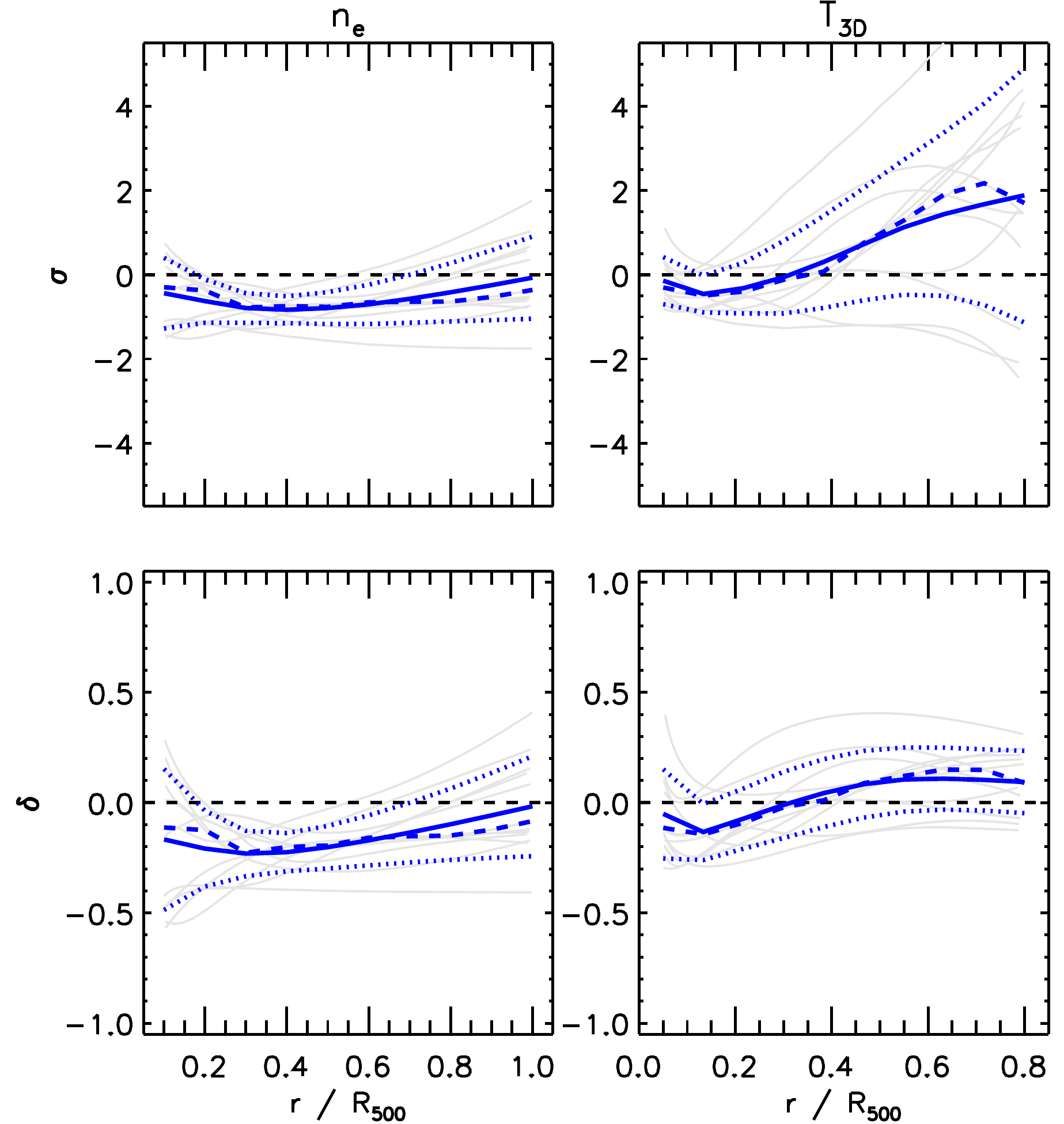} 
\caption{Differences between the predicted and observed gas density and temperature profiles using the datasets from S18 (four panels on the left) and G19 (four panels on the right). 
(Top panels) Deviations in terms of $\sigma = (O-M) / \epsilon$, where $O$ is the observed profile with error $\epsilon_O$ and $M$ is the predicted value with an estimated uncertainty of $\epsilon_M$ (see Fig.~\ref{fig:syst}) that combines in quadrature with $\epsilon_O$ to produce the total error $\epsilon$; and (bottom panels) systematic deviation $\delta = (O-M) / M$.
Grey lines represent the single object; solid, dashed, and dotted lines indicate mean, median, and scatter, respectively. 
In the four panels on the left, blue and red lines correspond to the properties of the subsamples at $z<0.6$ and $z>0.6$, respectively.
} \label{fig:s18g19dat}
\end{figure*}

\section{Radial profiles: Calibration in mass and redshift}
\label{sect:radprof}

Characteristic thermodynamic quantities defined within an overdensity $\Delta$, $Q_{\Delta}$, were used to renormalize the radial profiles of the given quantity only as a function of the halo properties (i.e., mass and redshift; see Tab.~\ref{tab:normz}). 
Effects of the deviation from self-similarity propagate to these values in a way that can be represented in the form 
\begin{equation}
    Q_{\Delta} \sim \, M^{a_M} \, E_z^{a_z} \sim \, T^{a_T} \, E_z^{a_{Tz}}.
\label{eq:qdelta}
\end{equation}
Self-similar and modified values, after the calibration in the context of the \icmz\ model presented in Eq.~\ref{eq:f1t1} and \ref{eq:fztz}, are quoted in Table~\ref{tab:normz_bestfit}.

Recently, \cite{pratt22} have estimated the evolution and mass dependence of the ``universal'' gas density profile for a sample of 93 SZ-selected galaxy clusters in the mass range $M_{500} = [0.5 - 20] \times 10^{14} M_{\odot}$ at $0.05<z<1.13$.
They measured $\bar{a}_{M, n} = 0.22 \pm 0.01$ and $\bar{a}_{z, n} = 2.09 \pm 0.02$.
Using the best-fit values in equations~\ref{eq:f1t1} and \ref{eq:fztz}, we obtained with the \icmz\ model $a_{M, n} = 0.23 (\pm 0.01)$ and $a_{z, n} = 2.11 (\pm 0.03)$, which are in remarkable agreement with both the dependencies.

A constraint of the universal pressure profile has been presented in \cite{arnaud10}: $\bar{a}_{M, P} = 2/3 +a_P +a_{P1} \approx 0.88$ with $a_P=0.12$ and $a_{P1} = 0.10 -(a_P+0.10) \times (x/0.5)^3/(1+(x/0.5)^3)$ and $x\approx0.1$, where a self-similar evolution of $\bar{a}_{z, P} = 8/3 = 2.67$ has been assumed.
We obtained $a_{M, P} = 0.90 (\pm 0.01)$ and $a_{z, P} = 2.77 (\pm 0.03)$, which does indeed suggest a deviation in the redshift dependence from the self-similar model as well.
These predicted values of $a_{M, P}$ and $a_{z, P}$ are consistent with the looser constraints we obtained by fitting the gas density and temperature profiles directly from \cite{sanders18} and \cite{ghi19univ} (see Fig.~\ref{fig:s18g19}).

\cite{pratt+10} constrained the mass dependence of the entropy profiles at a given overdensity. For $\Delta=500$, they measured $\bar{a}_{M, K} = 0.62 \pm 0.17$, still assuming a self-similar behavior in redshift ($\bar{a}_{z, K} = -2/3 = -0.67$).
With the calibrated \icmz\ model, we predict $a_{M, K} = 0.51 (\pm 0.01)$ and $a_{z, K} = -0.74 (\pm 0.02)$.

The predicted and observed profiles were compared and the free parameters ($a_m, a_z$) were constrained through a minimization of the following $\chi^2_m$:
\begin{equation}
    \chi^2_m = \sum_i \left[ \frac{ \left( d_i - m_i \right)^2 }{\epsilon_{d_i}^2 +\sigma_{m_i}^2} +\ln{\left( \epsilon_{d_i}^2 +\sigma_{m_i}^2 \right)} \right],
    \label{eq:chi2}
\end{equation}
where the index $m$ refers to the observable under investigation (e.g., surface brightness, $k=x$; gas density, $k=n$; and gas temperature, $k=T$); $m_i$ are the predicted profiles for the given observable from \icmz\ estimated for each object $i$ for the assumed mass and redshift; and $d_i$ and $\epsilon_{d_i}$ are the values, and the relative error, of the investigated quantities.
The quantity $\sigma_{m_i}$ accounts for the systematic uncertainties affecting the adopted model $m_i$. We have estimated those through a Monte-Carlo process where we propagated the errors on the assumed mass dependence and the scatter on both the $c-M-z$ relation and on the universal pressure profiles. 
For the latter ones, we adopted values of 0.16 in $\log_{10} c$ for a given mass \citep[e.g.,][]{dk15}, and 0.10 in $\log_{10} P$ \citep[e.g.,][]{arnaud10}.
We show in Fig.~\ref{fig:syst} an example of the estimates of $\sigma_{m_i}$ we recovered. 
For the sake of simplicity, we propagated the median relative errors to eq.~\ref{eq:chi2}.

Depending on the data available, more methods can be combined by adding the corresponding $\chi_m^2$.
For instance, when the profiles of the gas density and spectral temperature are in use, the total $\chi^2$ is estimated as $\chi_n^2 +\chi_T^2$.

We compared the predictions from the model \icmz\ with the radial profiles available in literature for a few recent datasets:

\begin{description}
\item[{\bf \cite{sanders18} (S18)}] The sample of \cite{sanders18} includes 83 galaxy clusters detected at the South Pole Telescope for their SZ signal and studied in X-ray with \cxo\ exposures. They span a range in $M_{500}$ between $1.2$ and $17.6 \times 10^{14} M_{\odot}$ at redshift $0.28-1.22$, with a median value of $4.6 \times 10^{14} M_{\odot}$ and $z=0.6$. The authors recovered the thermodynamic profiles through a publicly available forward-modeling projection code, MBPROJ2, that combines the information from exposure-corrected and background-subtracted X-ray surface brightness profiles extracted in ten independent energy bands between 0.5 and 7 keV, assuming hydrostatic equilibrium.
We used the publicly available density and temperature profiles.

\item[{\bf \cite{ghi19univ} (G19)}] This work describes the universal profiles recovered for the 12 massive ($M_{500} > 3 \times 10^{14} M_{\odot}$; median value, $5.7 \times 10^{14} M_{\odot}$), local ($0.05<<0.1$; median value, $0.065$) clusters constituting the \xmms\ Cluster Outskirts Project (X-COP) sample \citep{eck17xcop}, that is a very large program (VLP) on \xmm\ that targeted 12 the most significant Sunyaev-Zel’dovich (SZ) sources in the Planck survey in order to combine X-ray and SZ information out to the virial radius.
\end{description}

In Fig.~\ref{fig:s18g19dat}, we present the reconstructed profiles for the S18 and G19 samples. The deviations in $\sigma$ are, on average, within 1 over the radial range $0-1 \; R_{500}$ both in gas density and in temperature, with no evident dependency on the redshift. For the gas density, a systematic mean deviation $\delta$ from --30\% to +30\% moving outward in S18, and of about --10\% in G19 is observed; for the temperature profile, $\sim$ +20/30\% ($<$10\%) was measured in S18 (G19), with the scatter around these central values, however, encompassing the zero corresponding to the perfect match.

\section{Integrated quantities: Predictions and comparison with the modified scaling laws}
\label{sect:glob}

\begin{figure}[ht]
\includegraphics[trim=50 40 60 220,clip,width=\hsize]{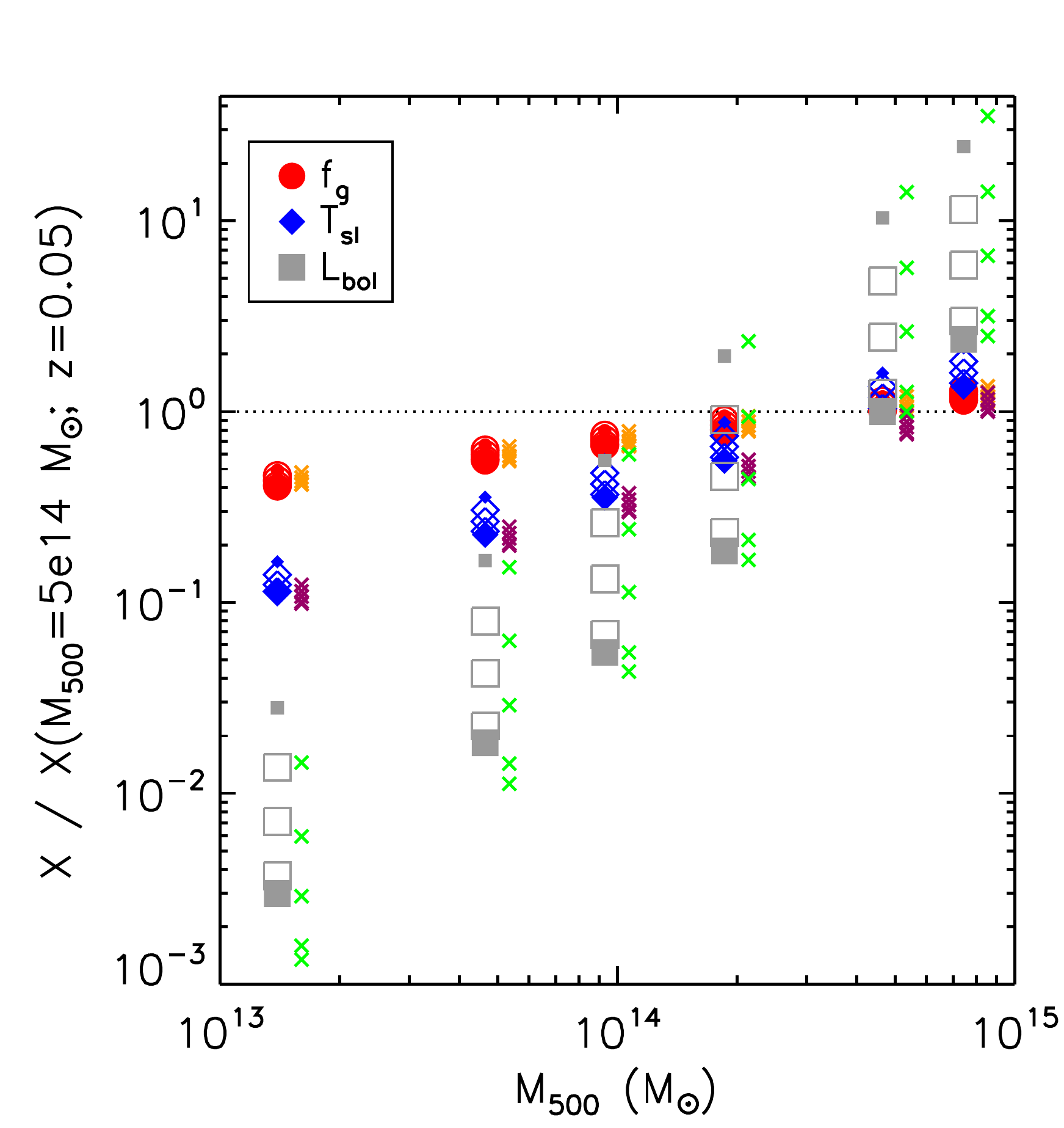}  
\caption{Comparison between some interesting global values $X$ (hydrostatic mass, spectroscopic-like temperature, bolometric luminosity, and gas mass fraction) estimated from the \icmz\ model and the predicted revised scaling laws (see equations~\ref{eq:scalaw_ez}). 
For each quantity $X$, we plot the value normalized at $(M_{500}, z) = (5\times10^{14} M_{\odot}, 0.05)$ and estimated at $\Delta=500$ estimated for the grids of $M_{500}/ 10^{14} M_{\odot} = [0.15, 0.5, 1, 2, 5, 8]$ and redshift $[0.05, 0.2, 0.6, 1, 1.5]$.
Filled dots indicate the estimates at the extreme of the redshift distribution, with the smallest dots corresponding to the highest $z$.
The crosses show the predicted values from the revised scaling laws (orange, $f_g$; purple, $T_{\rm sl}$; and green, $L_{\rm bol}$).
} \label{fig:glob}
\end{figure}

As detailed in E20, the \icmz\ model provides an estimate of the global quantities by integrating the thermodynamic profiles over the volume of interest.
In Fig.~\ref{fig:glob}, we compare these estimates for a set of representative quantities (gas mass fraction, gas temperature, and luminosity) with the relations that should stand among them (see Sect.~\ref{sect:icmz} and equations~\ref{eq:scalaw_ez}) once the revised laws have been calibrated as described in Sect.~\ref{sect:calib}.

In general, the agreement is remarkable; for instance, the gas mass fraction was recovered within 1\% over the entire range of mass $(10^{13}-10^{15} M_{\odot})$ and redshift $(0-1.5)$  investigated here.\ In addition, the spectroscopic temperature matches the predictions from the modified scaling laws, for a given mass and redshift, within 1\% (median value; with the range of the first and third quartile being --7\%  and 9\%).
Larger tension between the estimated and predicted values is present on the (bolometric) luminosity: the median (mean) differences are of 8\% (26\%), with the largest deviations (>50\%) associated with the haloes at $M_{500} < 5 \times10^{13} M_{\odot}$ (at higher masses, the median deviation is --4\% with first and third quartile of --13\% and 9\%).
We obtained a better agreement by considering the X-ray luminosity in a soft band (e.g., 0.5--2 keV), with a the median deviation of about 5\% (first and third quartile of  0 and 15\%).

\section{Constraints on the hydrostatic bias}
\label{sect:bhe}

\begin{figure}[htb]
\includegraphics[trim=20 30 40 400,clip,width=\hsize]{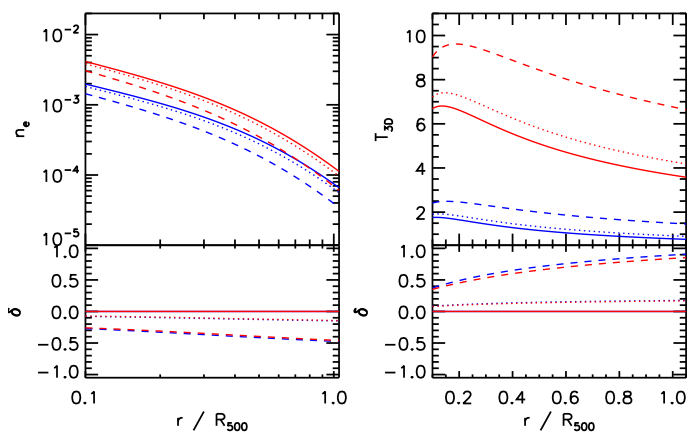} 
\caption{Predicted electron density (left) and temperature (right) profiles of two halos at $z=0.05$ and a hydrostatic mass $M_{500}$ of $5 \times 10^{13} M_{\odot}$ (blue lines) and $5 \times 10^{14} M_{\odot}$ (red lines), respectively.
Different line styles indicate the level of hydrostatic bias: $b=0$ (solid lines); $b=0.2$ (dotted line); and $b=0.6$ (dashed line).
Bottom panels: differences $\delta = (M_b - M_0)/M_0$ between the profiles with respect to the case with $b=0$.
} \label{fig:bhe}
\end{figure}

The procedure described in Sect.~\ref{sect:icmz} allows the hydrostatic and the total mass to be treated separately, with the consequent possibility of being to constrain the hydrostatic bias $b=1-M_{\rm HE}/M$. 
Overall, the method is sensitive to two effects: (i) the rescaling of the radius with $R_{500}$, with the latter being an estimate derived from the hydrostatic mass $M_{\rm HE} \equiv M_{500}$ defined as input; and (ii) the actual depth of the potential well, and thus the measurement of the gas temperature.
We show in Fig.~\ref{fig:bhe} the predicted impact of different levels of $b$:
with no strong dependence on the halo mass, the gas density should vary by about 10-15\% in the radial range $0.1-1 R_{500}$ for $b=0.2$, and between 30\% and 50\% for $b=0.6$; the gas temperature should be higher by $\sim$ 10\% when $b=0.2$ and up to 90\% with respect to the expected value evaluated with no bias when $b=0.6$.
To constrain the hydrostatic bias, the technique looks for a minimum $\chi^2$ for a grid of values of $b$ adopted as input, together with the other parameters of the \icmz\ model, the hydrostatic mass $M_{500}$, and the redshift. 

\begin{figure}[ht]
\includegraphics[trim=50 50 70 230,clip,width=\hsize]{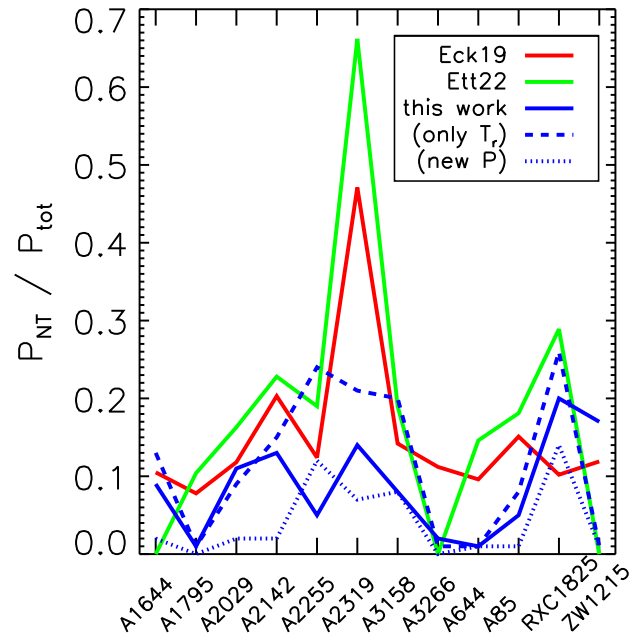} 
\caption{Predicted $1 \sigma$ upper limits on the value of the hydrostatic bias $b$ here represented as the ratio between the nonthermal ($P_{\rm NT}$) and total ($P_{\rm tot}$) pressure in the ICM \citep[$b = P_{\rm NT}/P_{\rm tot}$; see e.g.,][]{eckert19,ettori22}.
The dashed line represents the upper limit when only the temperature profile was used in the likelihood; when the new universal pressure profile also formed X-COP data (see Sect.~\ref{sect:upp}), the upper limits indicated with a dotted line were obtained.
We note that the high level of nonthermal pressure support required for A2319 is explained by the ongoing merger described in \cite{ghi18}.
} \label{fig:bxcop}
\end{figure}

We have applied this technique to the X-COP sample, for which independent measurements of $b$ have been obtained either by comparing different mass estimators \citep[][]{ettori19}, or by imposing a universal gas mass fraction and various models of nonthermal pressure support \citep[][]{eckert19,ettori22}.
We show the best-fit constraints for all 12 objects in Fig.~\ref{fig:xcop}.
Table~\ref{tab:bxcop} and Fig.~\ref{fig:bxcop} summarize the limits from the published and current work.
The technique proposed in this study does indeed seem to identify the objects more affected from a hydrostatic bias; 
although, they do suggest values on the lower end of the distribution overall.

\begin{table}[ht]
\caption{Hydrostatic bias $b (R_{500})$ in X-COP clusters.} 
\setlength\tabcolsep{2.2pt} \begin{center} \begin{tabular}{ccccccc} \hline
Cluster & $\alpha_{500}$ (Eck19) & $\alpha_{500}$ (Ett22) & $b$ (this work) / only $\chi^2_T$ \\ \hline
A1644 & $<0.10$ & $<0.00$ & $0.05_{-0.04}^{+0.04}$ /  $0.09_{-0.05}^{+0.04}$ \\
A1795 & $0.02_{-0.02}^{+0.06}$ & $0.07_{-0.02}^{+0.04}$ & $<0.01$ /  $<0.01$ \\
A2029 & $0.06_{-0.06}^{+0.06}$ & $0.11_{-0.04}^{+0.05}$ & $0.06_{-0.05}^{+0.05}$ /  $0.03_{-0.03}^{+0.06}$ \\
A2142 & $0.16_{-0.05}^{+0.05}$ & $0.20_{-0.02}^{+0.03}$ & $0.09_{-0.03}^{+0.04}$ /  $0.11_{-0.05}^{+0.04}$ \\
A2255 & $0.06_{-0.06}^{+0.07}$ & $0.11_{-0.02}^{+0.08}$ & $<0.05$ /  $0.18_{-0.07}^{+0.06}$ \\
A2319 & $0.44_{-0.04}^{+0.04}$ & $0.54_{-0.07}^{+0.12}$ & $0.12_{-0.02}^{+0.02}$ /  $0.19_{-0.02}^{+0.02}$ \\
A3158 & $0.09_{-0.06}^{+0.06}$ & $0.13_{-0.06}^{+0.06}$ & $0.04_{-0.04}^{+0.04}$ /  $0.15_{-0.06}^{+0.05}$ \\
A3266 & $<0.11$ & $<0.00$ & $<0.02$ /  $<0.01$ \\
A644 & $0.03_{-0.03}^{+0.06}$ & $<0.15$ & $<0.01$ /  $<0.01$ \\
A85 & $0.10_{-0.06}^{+0.05}$ & $0.15_{-0.03}^{+0.03}$ & $0.01_{-0.01}^{+0.04}$ /  $0.04_{-0.04}^{+0.04}$ \\
RXC1825 & $0.05_{-0.05}^{+0.05}$ & $<0.29$ & $0.17_{-0.03}^{+0.03}$ /  $0.22_{-0.03}^{+0.04}$ \\
ZW1215 & $<0.12$ & $<0.00$ & $0.11_{-0.06}^{+0.06}$ /  $<0.01$ \\
\hline \end{tabular} \end{center}
\label{tab:bxcop}
\end{table}

We further considered the role of the universal pressure profile, in particular, by estimating a profile through a joint fit of the X-COP nonparametric deprojected points from the recent analysis in \cite{eckert22a} and including the corrections due to the nonthermal pressure to $M_{500}$ (and, consequently, to $R_{500}$) from the analysis in \cite{eckert19} (see Sect.~\ref{sect:upp} for details).
The purpose of this exercise was to evaluate how the universal pressure profile changes once the nonthermal pressure contribution is included and propagated to the rescaling in the radius and normalization. 
The new radial profile is latter than the one adopted here from \cite{arnaud10} (see Fig.~\ref{fig:upp_cf}), implying, for a given mass, lower (higher) values of gas density (temperature). 
The overall net effect is the reduction of the estimated mass bias. 

A larger effort as to the calibration of the universal pressure profile, also relying on careful and possibly bias-corrected mass estimates, is needed to strengthen the constraints through this method. 
Reversely, the method outlined here can potentially be used to assess the robustness of the assumption on the universal gas fraction that is the imposed to constrain the level of nonthermal pressure in
\cite{eckert19} and \cite{ettori22}.


\section{Limitations of the \icmz\ model}
\label{sect:limits}

The \icmz\ model, introduced in E20 and extended in the present work, relies on  the following:
(i) some relations between the total mass and halo concentration and redshift as well as a universal pressure profile put in hydrostatic equilibrium, and (ii) some simple modifications of the scaling relations that account from the departure from the self-similar predictions (see Equations~\ref{eq:scalaw} and \ref{eq:scalaw_ez}) also for what concerns the normalizations of the universal profiles (see Tab.~\ref{tab:normz}).
In E20, we discuss what the impact is of assuming alternative $c-M-z$ relations and some different sets of parameters describing the universal pressure profile.
The \icmz\ model has been calibrated here with some relations among X-ray-integrated quantities published in recent work.
As we discuss in Sect.~\ref{sect:calib}, each analyzed sample provides constraints on the parameters of the model (primarily $f_1$ and $t_1$, and then $f_z$ and  $t_z$) that differ between them and that might depend on the relations used and on the characteristic of the sample selected.
With the purpose of providing a first, general calibration of our model, we relied on all the scaling laws published recently for samples of hundreds of objects over a large range of masses and redshifts.

Some quantities are more prone than others to the few assumptions made in reconstructing them.
For instance, the X-ray luminosity depends on the assumed metallicity \citep[0.3 times the solar value as tabulated in][]{aspl09}. We have also investigated what the impact is in assuming a different value for the mean metallicity or a different table of reference. 
Using a metallicity of 1 solar, the tension between the predicted and recovered luminosity presented in Fig.~\ref{fig:glob} increases by more than a factor of 4.
On the contrary, still assuming a value of 0.3, but with respect to the abundance table in \cite{ag89}, the median differences increase to about 15\% for the X-ray luminosity in 0.5--2 keV  (first and third quartile of  8 and 27\%) and to 17\% for the pseudo-bolometric one (0 and 66\%).
 This assumption has a more subtle impact on the exponent of the cooling function \citep[see also e.g.,][]{lov21}. While the assumption on the solar table does not significantly affect (at fixed metallicity) the slope ($c \approx 0.43-0.44$ in the pseudo-bolometric band 0.01-100 keV; $c \approx -0.1$ in the soft X-ray bands); changes in metallicity have a larger impact, with $c$ decreasing to about $0.37$ and $-0.18$ for the pseudo-bolometric and soft bands, respectively, for a metal abundance of 1 solar, which induce a small change (of about 10\%) on the estimate of $f_1$.

Furthermore, we decided to calibrate the only few parameters that keep the model simple and, at the same time and more importantly, effective. A possible extension of the model would be the inclusion of a parameter representing the dynamical state \citep[i.e., relaxed and disturbed, see e.g.,][]{campitiello22}, which, for instance, might affect the choice of the $c-M-z$ relation and of the universal pressure profile, both showing slight differences between the most relaxed systems and the majority of the population.

\section{Summary and conclusions}

We present a detailed calibration in temperature (mass) and redshift of the \icmz\ model we have introduced in \cite{ettori20}.
The calibration was done using both a large sample of recently evaluated scaling relations \citep[see][]{lovisari20,bahar22,pratt22} and spatially resolved thermodynamic quantities \citep[from][]{sanders18,ghi19univ} 
at halo masses down to $M_{500} \approx 10^{13} M_{\odot}$ and up to redshift 1.2. 

Our main findings are as follows.
   \begin{enumerate}[(i)]
      \item We effectively modified the self-similar scenario by introducing temperature- (mass-)dependent quantities such as the gas mass fraction $f_g$, the variation between its value at $R_{500}$, and the global spectroscopic temperature $T_{\rm spec}$, $f_T = T(R_{500}) / T_{500} \times T_{500} / T_{\rm spec} = T(R_{500}) / T_{\rm spec}$; by expressing the dependencies on the gas temperature and redshift as $f_g \sim T^{f_1} E_z^{f_z}$ and $f_T \sim T^{t_1} E_z^{t_z}$, we recovered how the slope and redshift evolution of the (mostly X-ray) scaling relations have to be modified (see Table~\ref{tab:normz}, and Sections~\ref{sect:SSmod} and \ref{sect:zevol}).
      \item Using a large dataset of published slopes and redshift evolution of X-ray scaling laws, in Sect.\ 3, we have constrained $\bar{f_1} = 0.403 \; (\pm 0.009)$, $\bar{t_1} = 0.144 \; (\pm 0.017)$, $\bar{f_z} = -0.004 \; (\pm 0.023)$, and $\bar{t_z} = 0.349 \; (\pm 0.059)$. These constraints are represented as predictions of the values of $f_g$ and $f_T$ as a function of the halo mass and redshift in Fig.~\ref{fig:ftfg}.\ We expect that the gas fraction $f_g$ does not change with the cosmic time, whereas a mild positive evolution is predicted for $f_T$; on the contrary, $f_g$ decreases significantly with $M_{500}$ (by 50\% going down in mass by an order of magnitude), and definitely more drastically than $f_T$.
      \item We verified that these values are well in agreement with the (looser) constraints provided by the observed thermodynamic profiles of large samples of galaxy clusters \citep{sanders18,ghi19univ}.
      \item By propagating the modified laws that account for a dependency on both the mass and the redshift, the \icmz\ model also allows one to modify any ``virial'' quantity accordingly and to evaluate how it affects the rescaling of any universal profile (see Sect.~\ref{sect:radprof}).\ We used our best-fit values in equations~\ref{eq:f1t1} and \ref{eq:fztz} in the scaling relations detailed in Table~\ref{tab:normz}, and quote in Table~\ref{tab:normz_bestfit} the expected variations on the slope and redshift dependence of a given quantity $Q$ estimated within an overdensity $\Delta$ (see eq.~\ref{eq:qdelta}), finding remarkable agreement with the current observational constraints.
      \item In Sect.~\ref{sect:glob}, we have verified the self-consistency of the \icmz\ model by evaluating the integrated quantities from the thermodynamic profiles (as detailed in E20). The integrated values of some representative and fundamental quantities, such as the gas mass fraction and global temperature, match the predictions obtained from the modified and calibrated scaling laws within a few percent.\ Only the bolometric luminosity shows differences up to 100\%, but only in low mass ($M_{500} < 5 \times 10^{13} M_{\odot}$) systems.\ When a luminosity in a soft X-ray band is considered, the tension in these objects decreases by more than a factor of two; part of this tension on the recovered luminosity can be ascribed to the assumptions made as to the plasma metallicity (see discussion in Sect.~\ref{sect:limits}).
      \item In Sect.~\ref{sect:bhe}, we have validated the \icmz\ model as a reliable indicator of the level of hydrostatic bias present in the total mass reconstruction.\ By applying the technique to the X-COP objects, a sample that offers high-quality data from which a detailed analysis of their hydrostatic mass profile has been produced \citep{ettori19, eckert22a}, we assessed a level of hydrostatic bias quite in agreement with the current available constraints based on completely different assumptions \citep[e.g., an universal gas mass fraction of reference; see][]{eckert19, ettori22}.
   \end{enumerate}

\begin{figure}[ht]
\centering
\includegraphics[width=\hsize]{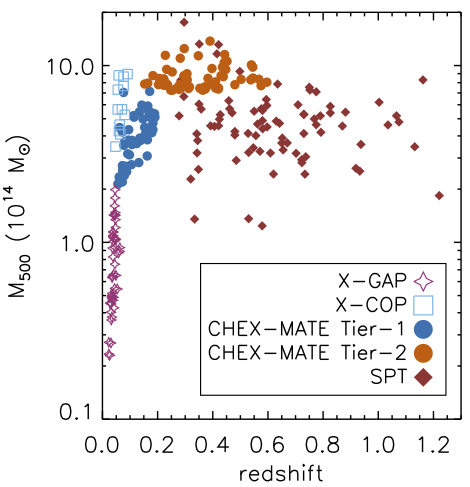} 
\caption{Distribution in the mass-redshift plane of the objects (X-COP, Ghirardini et al. 2019; SPT, Sanders et al. 2018) used in the present work for the calibration of the \icmz\ model, compared with the samples that will be available in the near future ({\it CHEX-MATE}, CHEX-MATE Collaboration et al. 2021; {\it X-GAP}).
} \label{fig:sample}
\end{figure}

Furthermore, we speculate that the \icmz\ model can be used to solve several variegated issues as follows.\ 
\begin{itemize}
    \item[$-$] A natural application is the construction of the expected observational quantities (from gas density to gas temperature and pressure profiles, from X-ray surface brightness profiles in an assigned observational band to the integrated X-ray luminosity, temperature, and gas mass) for each object in a given catalog of haloes with an assigned mass and redshift.
    \item[$-$] Its predictions can be used to validate and calibrate the outcomes of hydrodynamical simulations.
    \item[$-$] By fitting the observed thermodynamic radial profiles, it can provide an estimate of the relevance of the hydrostatic bias in a given object.
    \item[$-$] The calibrated model predicts that the gas mass fraction depends on the gas temperature, but it does not evolve with redshift, allowing for  the systematic uncertainties that affect the use of $f_{\rm gas} = M_{\rm gas} / M_{\rm tot}$ as a cosmological ruler to be defined formally \citep[see e.g.,][]{ettori+09,mantz22}.    \item[$-$] Once the thermodynamic profiles are calibrated, we predict that, for example, the effective polytropic index $\gamma = \partial \log P / \partial \log n$ in the range $0.8-1.2 R_{500}$ should decrease with increasing halo masses and at higher redshifts, spanning values within 5\% (f.i. from 1.27 in halos with $M_{500} = 10^{13} M_{\odot}$ at $z=0.05$ to 1.14 for unlikely objects with $M_{500} = 10^{15} M_{\odot}$ at $z=2$) of the canonical value of 1.2 \citep[see, e.g.,][]{ghi19poly}.    \item[$-$] The most extreme objects in their observed properties can be identified and their (eventual) tension quantified; for example, from equations~\ref{eq:scalaw} and \ref{eq:scalaw_ez} and the best-fit results in equations~\ref{eq:f1t1} and \ref{eq:fztz}, we predict that, in a $\Lambda$CDM universe, a galaxy cluster with a global $T$ of 5 keV for example is expected to be 4.5 times more luminous (bolometric luminosity; 3.5, if we refer to $L_{500}$ in the 0.5-2 keV band) than a 3 keV system, and 1.6, 2.3, and 5.0 times more luminous at $z=[0.6, 1, 2]$ than at $z=0.05$. Significantly higher (or lower) values, and outside the intrinsic scatter of the population investigated, would represent a conundrum for the cosmological formation of these systems.
\end{itemize}

Samples of spatially resolved thermodynamic properties out to $R_{500}$ at a high signal-to-noise ratio for a large number of objects will be soon available through numerous dedicated large \xmm\ programs, such as {\it CHEX-MATE}\footnote{\url{http://xmm-heritage.oas.inaf.it/}}, a Planck SZ selected sample of 118 objects covering masses higher than $M_{500} \sim 2 \times 10^{14} M_{\odot}$ up to $z\approx 0.6$ \citep{chexmate21} and {\it X-GAP}\footnote{\url{https://www.astro.unige.ch/xgap/}}, where 49 local galaxy groups with $M_{500}$ between $10^{13}$ and $10^{14} M_{\odot}$ will be observed homogeneously to measure $f_{\rm gas}(<R_{500})$ with statistical uncertainties of about 10\%  (see Fig.~\ref{fig:sample}).
For these data, we will expand our modelization by constraining the adopted parameters at higher
confidence, and introducing even new parameters (for instance related to the dynamical state), if needed.

\begin{acknowledgements}
We thank the anonymous referee for the positive and constructive comments.
S.E. and L.L. acknowledge financial contribution from the contracts ASI-INAF Athena 2019-27-HH.0, 
``Attivit\`a di Studio per la comunit\`a scientifica di Astrofisica delle Alte Energie e Fisica Astroparticellare'' (Accordo Attuativo ASI-INAF n. 2017-14-H.0), 
INAF mainstream project 1.05.01.86.10, and from the European Union’s Horizon 2020 Programme under the AHEAD2020 project (grant agreement n. 871158).
The \icmz\ model described in Sect.~\ref{sect:icmz} is encoded as an IDL function and available upon request from the first author. 
\end{acknowledgements}

\bibliographystyle{aa}
\bibliography{icmz}

\begin{thebibliography}{40}
\expandafter\ifx\csname natexlab\endcsname\relax\def\natexlab#1{#1}\fi

\bibitem[{{Allen} {et~al.}(2011){Allen}, {Evrard}, \& {Mantz}}]{allen11}
{Allen}, S.~W., {Evrard}, A.~E., \& {Mantz}, A.~B. 2011, \araa, 49, 409

\bibitem[{{Anders} \& {Grevesse}(1989)}]{ag89}
{Anders}, E. \& {Grevesse}, N. 1989, \gca, 53, 197

\bibitem[{{Andrade-Santos} {et~al.}(2017){Andrade-Santos}, {Jones}, {Forman},
  {Lovisari}, {Vikhlinin}, {van Weeren}, {Murray}, {Arnaud}, {Pratt},
  {D{\'e}mocl{\`e}s}, {Kraft}, {Mazzotta}, {B{\"o}hringer}, {Chon},
  {Giacintucci}, {Clarke}, {Borgani}, {David}, {Douspis}, {Pointecouteau},
  {Dahle}, {Brown}, {Aghanim}, \& {Rasia}}]{andrade17}
{Andrade-Santos}, F., {Jones}, C., {Forman}, W.~R., {et~al.} 2017, \apj, 843,
  76

\bibitem[{{Angelinelli} {et~al.}(2022){Angelinelli}, {Ettori}, {Dolag},
  {Vazza}, \& {Ragagnin}}]{angelinelli22}
{Angelinelli}, M., {Ettori}, S., {Dolag}, K., {Vazza}, F., \& {Ragagnin}, A.
  2022, \aap, 663, L6

\bibitem[{{Arnaud} {et~al.}(2010){Arnaud}, {Pratt}, {Piffaretti},
  {B{\"o}hringer}, {Croston}, \& {Pointecouteau}}]{arnaud10}
{Arnaud}, M., {Pratt}, G.~W., {Piffaretti}, R., {et~al.} 2010, \aap, 517, A92

\bibitem[{{Asplund} {et~al.}(2009){Asplund}, {Grevesse}, {Sauval}, \&
  {Scott}}]{aspl09}
{Asplund}, M., {Grevesse}, N., {Sauval}, A.~J., \& {Scott}, P. 2009, \araa, 47,
  481

\bibitem[{{Bahar} {et~al.}(2022){Bahar}, {Bulbul}, {Clerc}, {Ghirardini},
  {Liu}, {Nandra}, {Pacaud}, {Chiu}, {Comparat}, {Ider-Chitham}, {Klein},
  {Liu}, {Merloni}, {Migkas}, {Okabe}, {Ramos-Ceja}, {Reiprich}, {Sanders}, \&
  {Schrabback}}]{bahar22}
{Bahar}, Y.~E., {Bulbul}, E., {Clerc}, N., {et~al.} 2022, \aap, 661, A7

\bibitem[{{Bhattacharya} {et~al.}(2013){Bhattacharya}, {Habib}, {Heitmann}, \&
  {Vikhlinin}}]{bha13}
{Bhattacharya}, S., {Habib}, S., {Heitmann}, K., \& {Vikhlinin}, A. 2013, \apj,
  766, 32

\bibitem[{{Campitiello} {et~al.}(2022){Campitiello}, {Ettori}, {Lovisari},
  {Bartalucci}, {Eckert}, {Rasia}, {Rossetti}, {Gastaldello}, {Pratt},
  {Maughan}, {Pointecouteau}, {Sereno}, {Biffi}, {Borgani}, {De Luca}, {De
  Petris}, {Gaspari}, {Ghizzardi}, {Mazzotta}, \& {Molendi}}]{campitiello22}
{Campitiello}, M.~G., {Ettori}, S., {Lovisari}, L., {et~al.} 2022, \aap, 665,
  A117

\bibitem[{{CHEX-MATE Collaboration} {et~al.}(2021){CHEX-MATE Collaboration},
  {Arnaud}, {Ettori}, {Pratt}, {Rossetti}, {Eckert}, {Gastaldello}, {Gavazzi},
  {Kay}, {Lovisari}, {Maughan}, {Pointecouteau}, {Sereno}, {Bartalucci},
  {Bonafede}, {Bourdin}, {Cassano}, {Duffy}, {Iqbal}, {Maurogordato}, {Rasia},
  {Sayers}, {Andrade-Santos}, {Aussel}, {Barnes}, {Barrena}, {Borgani},
  {Burkutean}, {Clerc}, {Corasaniti}, {Cuillandre}, {De Grandi}, {De Petris},
  {Dolag}, {Donahue}, {Ferragamo}, {Gaspari}, {Ghizzardi}, {Gitti}, {Haines},
  {Jauzac}, {Johnston-Hollitt}, {Jones}, {K{\'e}ruzor{\'e}}, {Le Brun},
  {Mayet}, {Mazzotta}, {Melin}, {Molendi}, {Nonino}, {Okabe}, {Paltani},
  {Perotto}, {Pires}, {Radovich}, {Rubino-Martin}, {Salvati}, {Saro},
  {Sartoris}, {Schellenberger}, {Streblyanska}, {Tarr{\'\i}o}, {Tozzi},
  {Umetsu}, {van der Burg}, {Vazza}, {Venturi}, {Yepes}, \&
  {Zarattini}}]{chexmate21}
{CHEX-MATE Collaboration}, {Arnaud}, M., {Ettori}, S., {et~al.} 2021, \aap,
  650, A104

\bibitem[{{Diemer} \& {Kravtsov}(2015)}]{dk15}
{Diemer}, B. \& {Kravtsov}, A.~V. 2015, \apj, 799, 108

\bibitem[{{Dutton} \& {Macci{\`o}}(2014)}]{dutton14}
{Dutton}, A.~A. \& {Macci{\`o}}, A.~V. 2014, \mnras, 441, 3359

\bibitem[{{Eckert} {et~al.}(2017){Eckert}, {Ettori}, {Pointecouteau},
  {Molendi}, {Paltani}, \& {Tchernin}}]{eck17xcop}
{Eckert}, D., {Ettori}, S., {Pointecouteau}, E., {et~al.} 2017, Astronomische
  Nachrichten, 338, 293

\bibitem[{{Eckert} {et~al.}(2022){Eckert}, {Ettori}, {Pointecouteau}, {van der
  Burg}, \& {Loubser}}]{eckert22a}
{Eckert}, D., {Ettori}, S., {Pointecouteau}, E., {van der Burg}, R.~F.~J., \&
  {Loubser}, S.~I. 2022, \aap, 662, A123

\bibitem[{{Eckert} {et~al.}(2021){Eckert}, {Gaspari}, {Gastaldello}, {Le Brun},
  \& {O'Sullivan}}]{eck21}
{Eckert}, D., {Gaspari}, M., {Gastaldello}, F., {Le Brun}, A. M.~C., \&
  {O'Sullivan}, E. 2021, Universe, 7, 142

\bibitem[{{Eckert} {et~al.}(2019){Eckert}, {Ghirardini}, {Ettori}, {Rasia},
  {Biffi}, {Pointecouteau}, {Rossetti}, {Molendi}, {Vazza}, {Gastaldello},
  {Gaspari}, {De Grandi}, {Ghizzardi}, {Bourdin}, {Tchernin}, \&
  {Roncarelli}}]{eckert19}
{Eckert}, D., {Ghirardini}, V., {Ettori}, S., {et~al.} 2019, Astronomy and
  Astrophysics, 621, A40

\bibitem[{{Eckert} {et~al.}(2011){Eckert}, {Molendi}, \& {Paltani}}]{eckert11}
{Eckert}, D., {Molendi}, S., \& {Paltani}, S. 2011, \aap, 526, A79

\bibitem[{{Ettori}(2015)}]{ettori15}
{Ettori}, S. 2015, \mnras, 446, 2629

\bibitem[{{Ettori} \& {Eckert}(2022)}]{ettori22}
{Ettori}, S. \& {Eckert}, D. 2022, \aap, 657, L1

\bibitem[{{Ettori} {et~al.}(2019){Ettori}, {Ghirardini}, {Eckert},
  {Pointecouteau}, {Gastaldello}, {Sereno}, {Gaspari}, {Ghizzardi},
  {Roncarelli}, \& {Rossetti}}]{ettori19}
{Ettori}, S., {Ghirardini}, V., {Eckert}, D., {et~al.} 2019, Astronomy and
  Astrophysics, 621, A39

\bibitem[{{Ettori} {et~al.}(2020){Ettori}, {Lovisari}, \& {Sereno}}]{ettori20}
{Ettori}, S., {Lovisari}, L., \& {Sereno}, M. 2020, \aap, 644, A111

\bibitem[{{Ettori} {et~al.}(2009){Ettori}, {Morandi}, {Tozzi}, {Balestra},
  {Borgani}, {Rosati}, {Lovisari}, \& {Terenziani}}]{ettori+09}
{Ettori}, S., {Morandi}, A., {Tozzi}, P., {et~al.} 2009, \aap, 501, 61

\bibitem[{{Ghirardini} {et~al.}(2022){Ghirardini}, {Bahar}, {Bulbul}, {Liu},
  {Clerc}, {Pacaud}, {Comparat}, {Liu}, {Ramos-Ceja}, {Hoang}, {Ider-Chitham},
  {Klein}, {Merloni}, {Nandra}, {Ota}, {Predehl}, {Reiprich}, {Sanders}, \&
  {Schrabback}}]{ghi22}
{Ghirardini}, V., {Bahar}, Y.~E., {Bulbul}, E., {et~al.} 2022, \aap, 661, A12

\bibitem[{{Ghirardini} {et~al.}(2019{\natexlab{a}}){Ghirardini}, {Eckert},
  {Ettori}, {Pointecouteau}, {Molendi}, {Gaspari}, {Rossetti}, {De Grandi},
  {Roncarelli}, {Bourdin}, {Mazzotta}, {Rasia}, \& {Vazza}}]{ghi19univ}
{Ghirardini}, V., {Eckert}, D., {Ettori}, S., {et~al.} 2019{\natexlab{a}},
  Astronomy and Astrophysics, 621, A41

\bibitem[{{Ghirardini} {et~al.}(2019{\natexlab{b}}){Ghirardini}, {Ettori},
  {Eckert}, \& {Molendi}}]{ghi19poly}
{Ghirardini}, V., {Ettori}, S., {Eckert}, D., \& {Molendi}, S.
  2019{\natexlab{b}}, Astronomy and Astrophysics, 627, A19

\bibitem[{{Ghirardini} {et~al.}(2018){Ghirardini}, {Ettori}, {Eckert},
  {Molendi}, {Gastaldello}, {Pointecouteau}, {Hurier}, \& {Bourdin}}]{ghi18}
{Ghirardini}, V., {Ettori}, S., {Eckert}, D., {et~al.} 2018, \aap, 614, A7

\bibitem[{{Lovisari} {et~al.}(2021){Lovisari}, {Ettori}, {Gaspari}, \&
  {Giles}}]{lov21}
{Lovisari}, L., {Ettori}, S., {Gaspari}, M., \& {Giles}, P.~A. 2021, Universe,
  7, 139

\bibitem[{{Lovisari} {et~al.}(2020){Lovisari}, {Schellenberger}, {Sereno},
  {Ettori}, {Pratt}, {Forman}, {Jones}, {Andrade-Santos}, {Randall}, \&
  {Kraft}}]{lovisari20}
{Lovisari}, L., {Schellenberger}, G., {Sereno}, M., {et~al.} 2020, \apj, 892,
  102

\bibitem[{{Ludlow} {et~al.}(2016){Ludlow}, {Bose}, {Angulo}, {Wang},
  {Hellwing}, {Navarro}, {Cole}, \& {Frenk}}]{lud16}
{Ludlow}, A.~D., {Bose}, S., {Angulo}, R.~E., {et~al.} 2016, \mnras, 460, 1214

\bibitem[{{Mantz} {et~al.}(2022){Mantz}, {Morris}, {Allen}, {Canning},
  {Baumont}, {Benson}, {Bleem}, {Ehlert}, {Floyd}, {Herbonnet}, {Kelly},
  {Liang}, {von der Linden}, {McDonald}, {Rapetti}, {Schmidt}, {Werner}, \&
  {Wright}}]{mantz22}
{Mantz}, A.~B., {Morris}, R.~G., {Allen}, S.~W., {et~al.} 2022, \mnras, 510,
  131

\bibitem[{{Navarro} {et~al.}(1997){Navarro}, {Frenk}, \& {White}}]{nfw97}
{Navarro}, J.~F., {Frenk}, C.~S., \& {White}, S.~D.~M. 1997, \apj, 490, 493

\bibitem[{{Nurgaliev} {et~al.}(2017){Nurgaliev}, {McDonald}, {Benson}, {Bleem},
  {Bocquet}, {Forman}, {Garmire}, {Gupta}, {Hlavacek-Larrondo}, {Mohr},
  {Nagai}, {Rapetti}, {Stark}, {Stubbs}, \& {Vikhlinin}}]{nurgaliev17}
{Nurgaliev}, D., {McDonald}, M., {Benson}, B.~A., {et~al.} 2017, \apj, 841, 5

\bibitem[{{Oppenheimer} {et~al.}(2021){Oppenheimer}, {Babul}, {Bah{\'e}},
  {Butsky}, \& {McCarthy}}]{opp21}
{Oppenheimer}, B.~D., {Babul}, A., {Bah{\'e}}, Y., {Butsky}, I.~S., \&
  {McCarthy}, I.~G. 2021, Universe, 7, 209

\bibitem[{{Planck Collaboration} {et~al.}(2013){Planck Collaboration}, {Ade},
  {Aghanim}, {Arnaud}, {Ashdown}, {Atrio-Barandela}, {Aumont}, {Baccigalupi},
  {Balbi}, {Banday}, {Barreiro}, {Bartlett}, {Battaner}, {Benabed},
  {Beno{\^\i}t}, {Bernard}, {Bersanelli}, {Bhatia}, {Bikmaev}, {Bobin},
  {B{\"o}hringer}, {Bonaldi}, {Bond}, {Borgani}, {Borrill}, {Bouchet},
  {Bourdin}, {Brown}, {Burenin}, {Burigana}, {Cabella}, {Cardoso}, {Carvalho},
  {Castex}, {Catalano}, {Cay{\'o}n}, {Chamballu}, {Chiang}, {Chon},
  {Christensen}, {Churazov}, {Clements}, {Colafrancesco}, {Colombi}, {Colombo},
  {Comis}, {Coulais}, {Crill}, {Cuttaia}, {Da Silva}, {Dahle}, {Danese},
  {Davis}, {de Bernardis}, {de Gasperis}, {de Zotti}, {Delabrouille},
  {D{\'e}mocl{\`e}s}, {D{\'e}sert}, {Diego}, {Dolag}, {Dole}, {Donzelli},
  {Dor{\'e}}, {D{\"o}rl}, {Douspis}, {Dupac}, {Efstathiou}, {En{\ss}lin},
  {Eriksen}, {Finelli}, {Flores-Cacho}, {Forni}, {Fosalba}, {Frailis},
  {Franceschi}, {Frommert}, {Galeotta}, {Ganga}, {G{\'e}nova-Santos}, {Giard},
  {Giraud-H{\'e}raud}, {Gonz{\'a}lez-Nuevo}, {G{\'o}rski}, {Gregorio},
  {Gruppuso}, {Hansen}, {Harrison}, {Hempel}, {Henrot-Versill{\'e}},
  {Hern{\'a}ndez-Monteagudo}, {Herranz}, {Hildebrandt}, {Hivon}, {Hobson},
  {Holmes}, {Hurier}, {Jaffe}, {Jaffe}, {Jagemann}, {Jones}, {Juvela},
  {Keih{\"a}nen}, {Khamitov}, {Kisner}, {Kneissl}, {Knoche}, {Knox}, {Kunz},
  {Kurki-Suonio}, {Lagache}, {L{\"a}hteenm{\"a}ki}, {Lamarre}, {Lasenby},
  {Lawrence}, {Le Jeune}, {Leonardi}, {Liddle}, {Lilje}, {L{\'o}pez-Caniego},
  {Luzzi}, {Mac{\'\i}as-P{\'e}rez}, {Maino}, {Mandolesi}, {Maris}, {Marleau},
  {Marshall}, {Mart{\'\i}nez-Gonz{\'a}lez}, {Masi}, {Massardi}, {Matarrese},
  {Mazzotta}, {Mei}, {Melchiorri}, {Melin}, {Mendes}, {Mennella}, {Mitra},
  {Miville-Desch{\^e}nes}, {Moneti}, {Montier}, {Morgante}, {Mortlock},
  {Munshi}, {Murphy}, {Naselsky}, {Nati}, {Natoli}, {N{\o}rgaard-Nielsen},
  {Noviello}, {Novikov}, {Novikov}, {Osborne}, {Pajot}, {Paoletti}, {Pasian},
  {Patanchon}, {Perdereau}, {Perotto}, {Perrotta}, {Piacentini}, {Piat},
  {Pierpaoli}, {Piffaretti}, {Plaszczynski}, {Pointecouteau}, {Polenta},
  {Ponthieu}, {Popa}, {Poutanen}, {Pratt}, {Prunet}, {Puget}, {Rachen},
  {Reach}, {Rebolo}, {Reinecke}, {Remazeilles}, {Renault}, {Ricciardi},
  {Riller}, {Ristorcelli}, {Rocha}, {Roman}, {Rosset}, {Rossetti},
  {Rubi{\~n}o-Mart{\'\i}n}, {Rusholme}, {Sandri}, {Savini}, {Scott}, {Smoot},
  {Starck}, {Sudiwala}, {Sunyaev}, {Sutton}, {Suur-Uski}, {Sygnet}, {Tauber},
  {Terenzi}, {Toffolatti}, {Tomasi}, {Tristram}, {Tuovinen}, {Valenziano}, {Van
  Tent}, {Varis}, {Vielva}, {Villa}, {Vittorio}, {Wade}, {Wandelt}, {Welikala},
  {White}, {White}, {Yvon}, {Zacchei}, \& {Zonca}}]{planck13}
{Planck Collaboration}, {Ade}, P.~A.~R., {Aghanim}, N., {et~al.} 2013, \aap,
  550, A131

\bibitem[{{Pratt} {et~al.}(2022){Pratt}, {Arnaud}, {Maughan}, \&
  {Melin}}]{pratt22}
{Pratt}, G.~W., {Arnaud}, M., {Maughan}, B.~J., \& {Melin}, J.~B. 2022, \aap,
  665, A24

\bibitem[{{Pratt} {et~al.}(2010){Pratt}, {Arnaud}, {Piffaretti},
  {B{\"o}hringer}, {Ponman}, {Croston}, {Voit}, {Borgani}, \&
  {Bower}}]{pratt+10}
{Pratt}, G.~W., {Arnaud}, M., {Piffaretti}, R., {et~al.} 2010, \aap, 511, A85

\bibitem[{{Rossetti} {et~al.}(2016){Rossetti}, {Gastaldello}, {Ferioli},
  {Bersanelli}, {De Grandi}, {Eckert}, {Ghizzardi}, {Maino}, \&
  {Molendi}}]{rossetti16}
{Rossetti}, M., {Gastaldello}, F., {Ferioli}, G., {et~al.} 2016, \mnras, 457,
  4515

\bibitem[{{Sanders} {et~al.}(2018){Sanders}, {Fabian}, {Russell}, \&
  {Walker}}]{sanders18}
{Sanders}, J.~S., {Fabian}, A.~C., {Russell}, H.~R., \& {Walker}, S.~A. 2018,
  \mnras, 474, 1065

\bibitem[{{Sayers} {et~al.}(2022){Sayers}, {Mantz}, {Rasia}, {Allen}, {Cui},
  {Golwala}, {Morris}, \& {Wan}}]{sayers22}
{Sayers}, J., {Mantz}, A.~B., {Rasia}, E., {et~al.} 2022, arXiv e-prints,
  arXiv:2206.00091

\bibitem[{{Voit}(2005)}]{voit05}
{Voit}, G.~M. 2005, Reviews of Modern Physics, 77, 207

\end{thebibliography}

\begin{appendix}

\section{Direct constraints on $a_M$ and $a_z$ from thermodynamic profiles}

\begin{figure}[ht]
\includegraphics[trim=35 40 300 270,clip,width=\hsize]{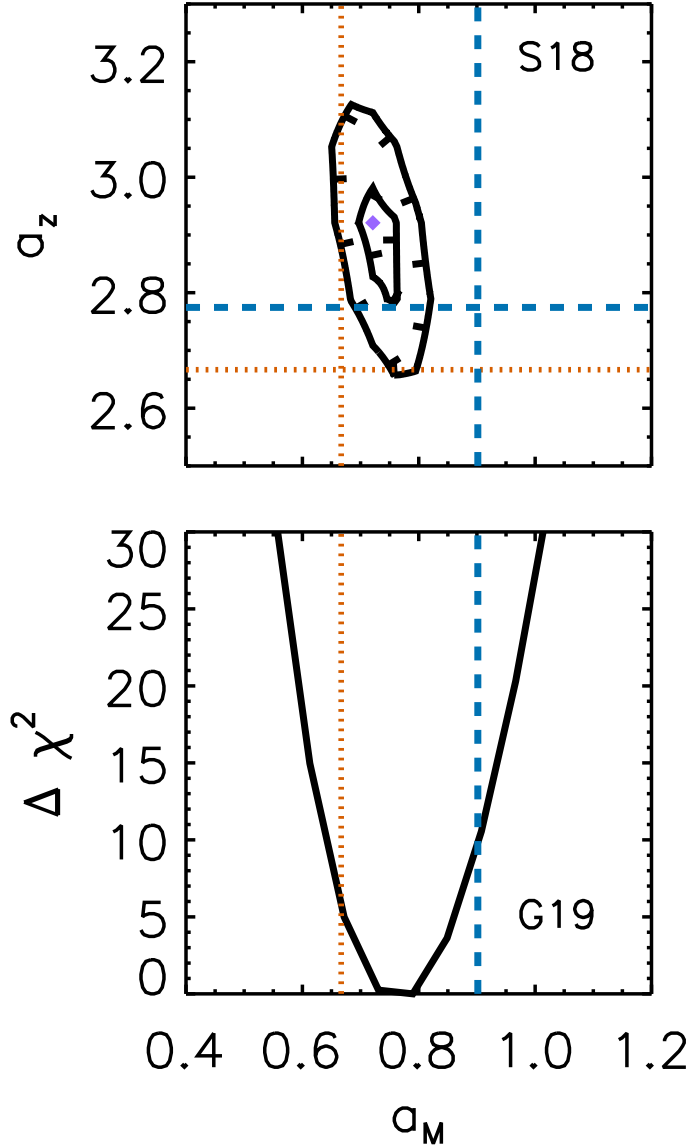} 
\caption{Constraints on the exponents $a_{M}$ and $a_{z}$ in $P_{\Delta} \sim M^{a_{M}} \; E_z^{a_{z}}$ based upon the gas density and temperature profiles in S18 (top panel; likelihood contour at $1$ and $5 \sigma$ equivalent to $\Delta \chi^2$ of 6.2 and 28.8 for 2 degrees of freedom, respectively) and G19 (bottom panel; on $a_M$ only, because all the 12 objects have a redshift $<0.1$. For the sake of comparison, $5 \sigma$ corresponds to $\Delta \chi^2$ of 25.0 for 1 degree of freedom). 
The dotted red lines are the predicted self-similar slopes, and the dashed blue lines correspond to the expected slopes from the \icmz\ model.
} \label{fig:s18g19}
\end{figure}

In Fig.~\ref{fig:s18g19}, we present the current constraints on $a_{M,P}$ and $a_{z,P}$ obtained by fitting the gas density and temperature profiles from the binned NFW fits (model BIN-NFW) in \cite{sanders18} and \cite{ghi19univ} shown in Fig.~\ref{fig:s18g19dat}. The gas density profiles were fitted over the radial range $0.15-1 R_{500}$, and the gas temperature profiles were fitted over the range $0.15-0.8 R_{500}$. 
A grid of $\chi^2$ values were estimated using Eq.~\ref{eq:chi2}, where the error budget on the model was obtained as described in Sect.~\ref{sect:radprof} for a representative object of $(M_{500}, z) = (4.6 \times 10^{14} M_{\odot}, 0.6)$ and $(6 \times 10^{14} M_{\odot}, 0.065)$ for S18 and G19, respectively.

\section{Hydrostatic mass bias corrected universal pressure profile in X-COP}
\label{sect:upp}

\begin{figure}[ht]
\includegraphics[trim=0 0 0 0,clip,width=\hsize]{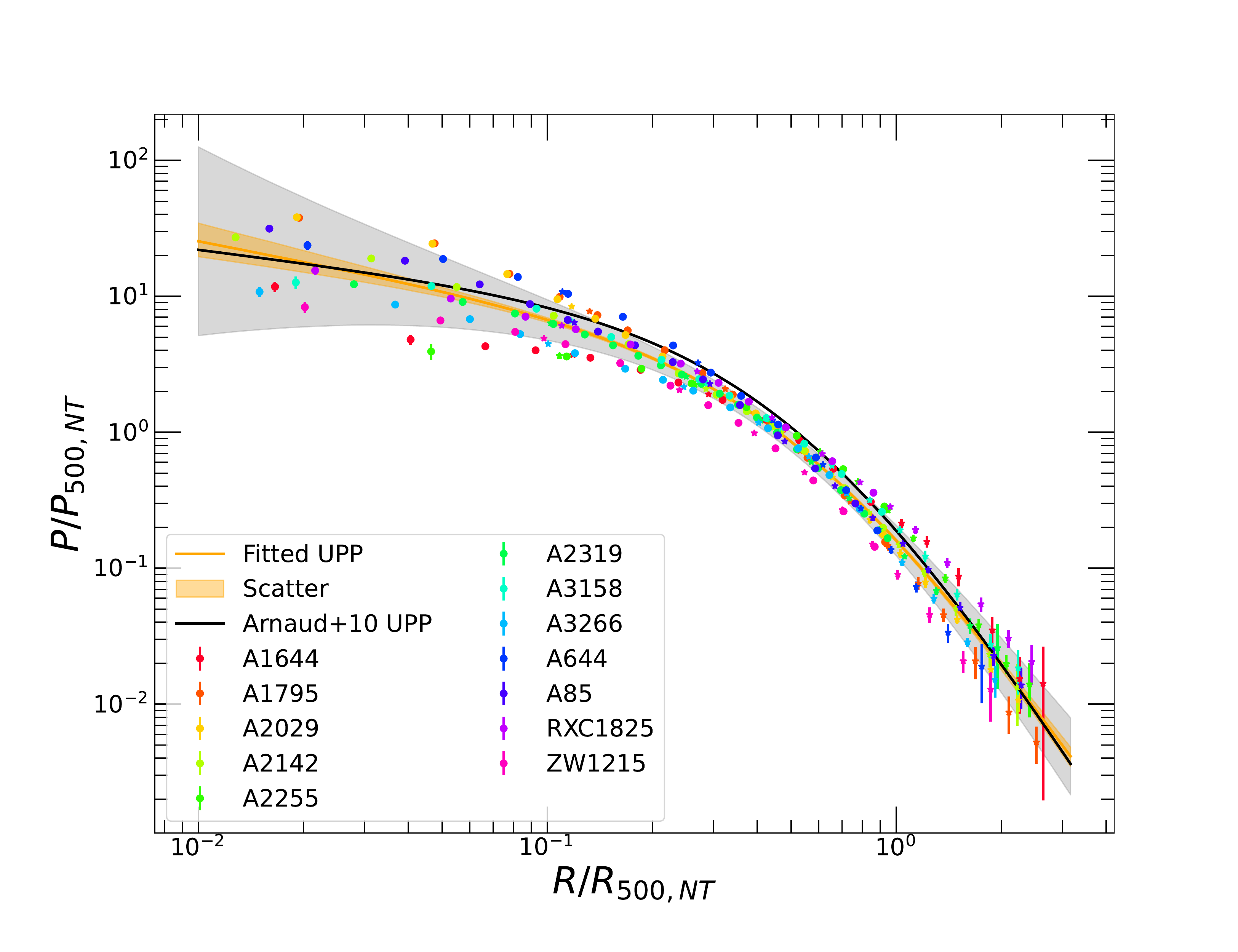} 
\caption{Rescaled pressure profiles for the X-COP objects after the analyses in \cite{eckert19, eckert22a}. We note that
$M_{500}$ and the relative $R_{500}$ were obtained after the correction for the estimated contribution from the nonthermal pressure \citep[see][for details]{eckert19}.
} \label{fig:upp_xcop}
\end{figure}

We have obtained a new estimate of the universal pressure profile (UPP) from the recent analysis of the X-COP data in \cite{eckert22a} by fitting
the nonparametric deprojected points rescaled with $M_{500}$ corrected by the nonthermal pressure contribution as described in \cite{eckert19}.
The rescaled profiles are shown in Fig.~\ref{fig:upp_xcop}.

\begin{figure}[ht]
\includegraphics[trim=0 0 0 0,clip,width=\hsize]{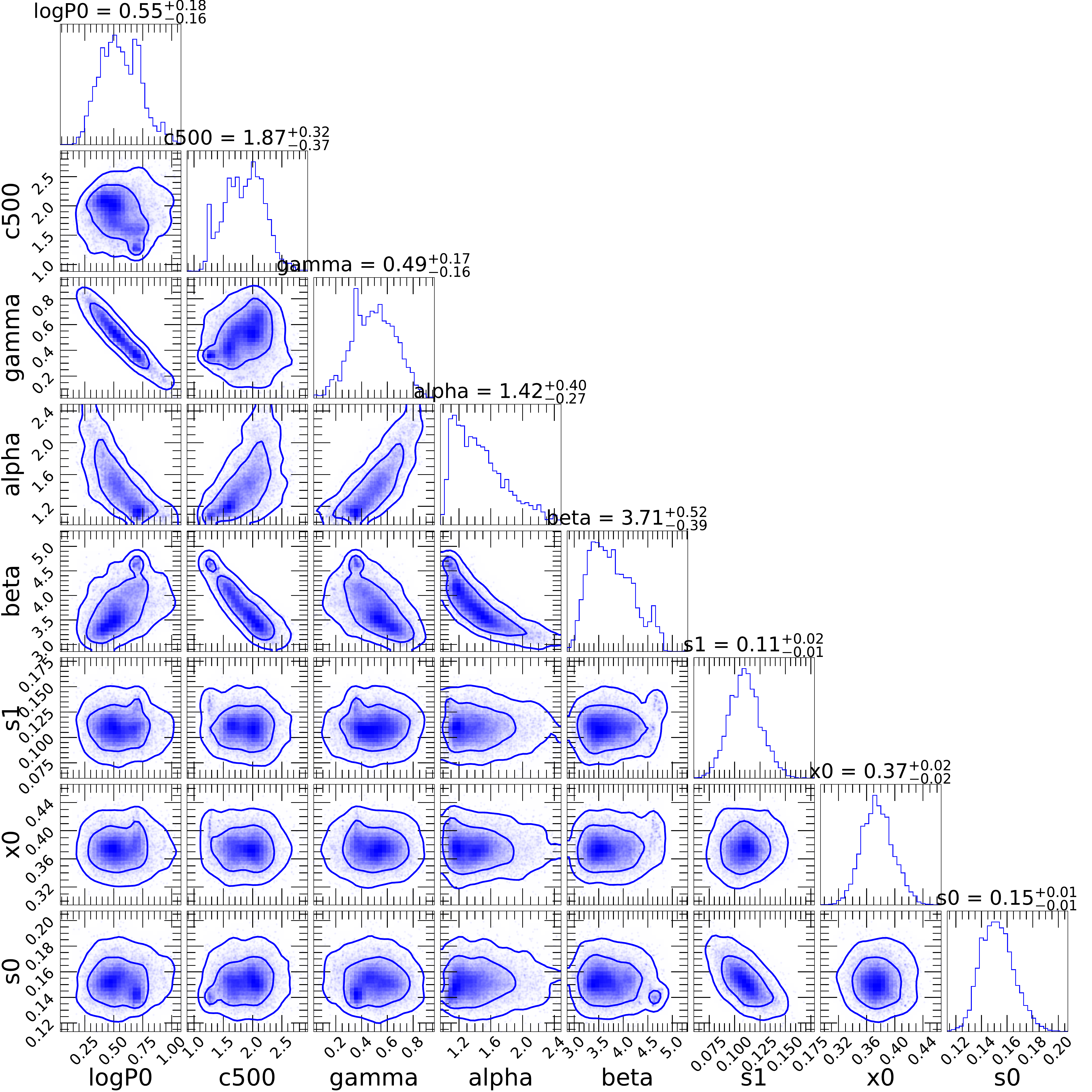}
\caption{Corner plot showing the constraints on the seven free parameters of the model in Eq.~\ref{eq:puniv}.}
\label{fig:upp_corner}
\end{figure}

\begin{figure}[ht]
\includegraphics[trim=0 0 0 0,clip,width=\hsize]{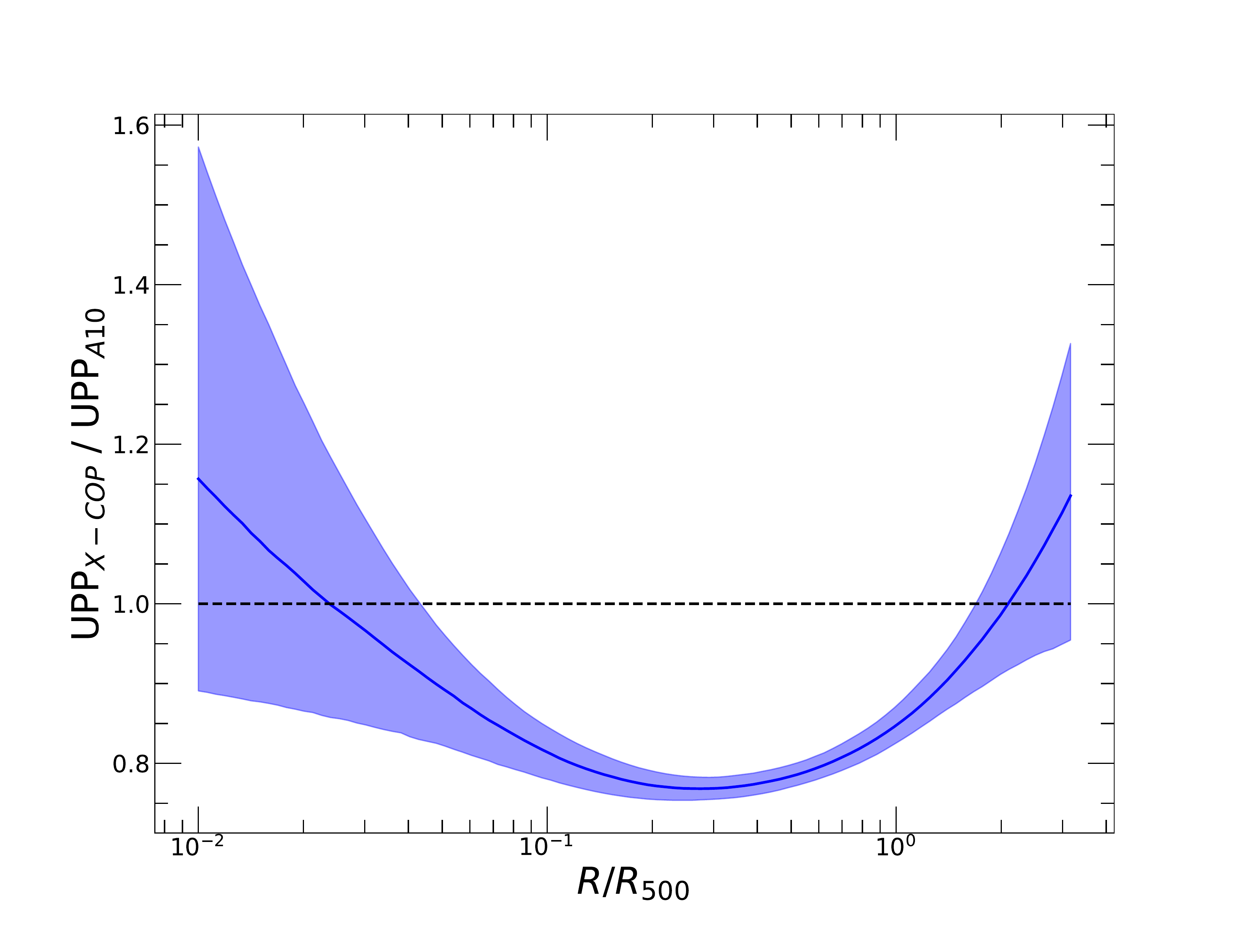}
\caption{Comparison between the universal pressure profiles (UPP) from the revised analysis of X-COP data and the one in \cite{arnaud10}.
} \label{fig:upp_cf}
\end{figure}

The scatter model is the log parabola from \cite{ghi19univ} (see their Eq. 6 and Fig.~7). The constraints on the free parameter of the model are shown in Fig.~\ref{fig:upp_corner}.
The profile differs from the UPP in \cite{arnaud10} only slightly, with a lower gas pressure at intermediate radii ($0.1-1 R_{500}$; see Fig.~\ref{fig:upp_cf}). 

\section{Hydrostatic bias in the X-COP}
\label{sect:xcop}

\begin{figure*}[ht]
\centering
\includegraphics[page=2,trim=0 40 0 550,clip,width=\hsize]{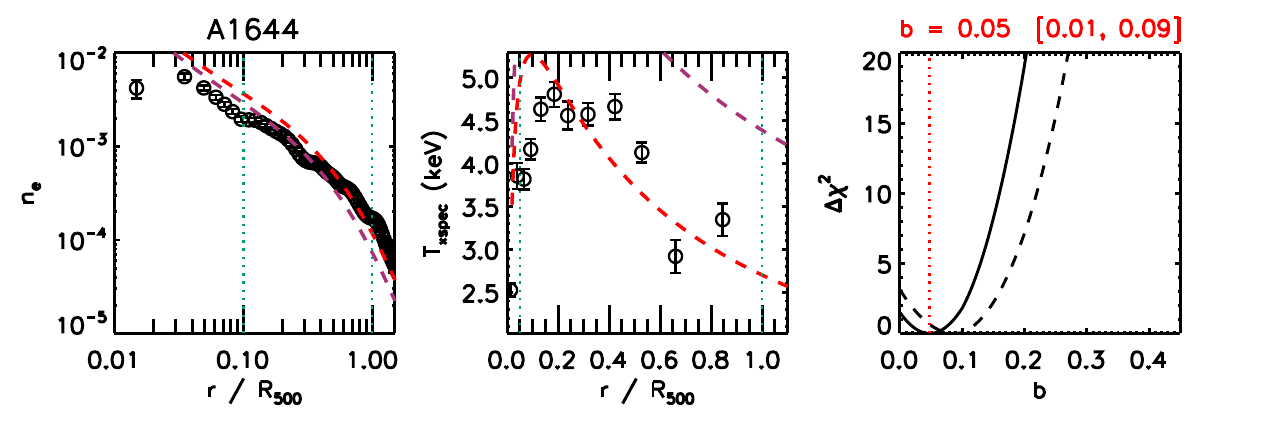} 
\includegraphics[page=4,trim=0 40 0 550,clip,width=\hsize]{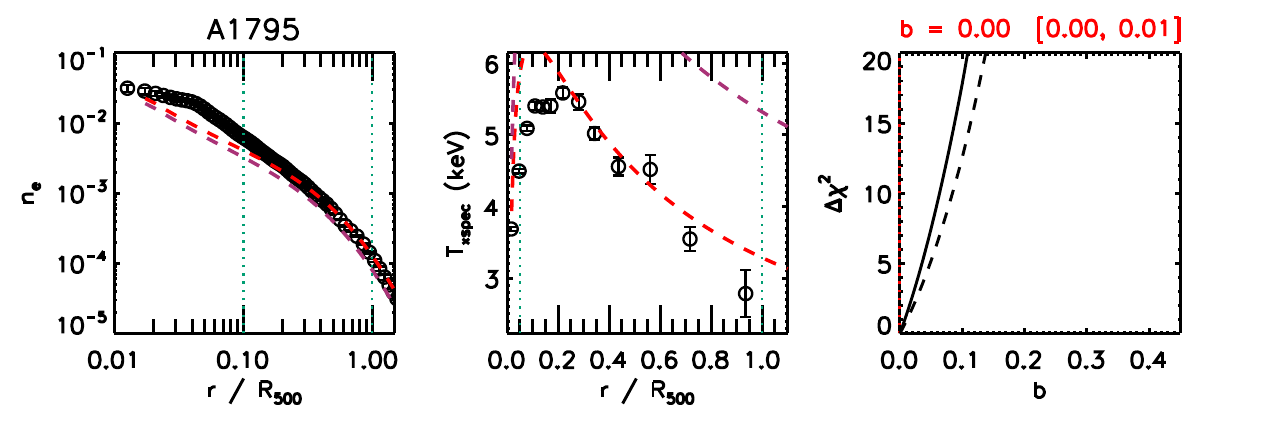} 
\includegraphics[page=6,trim=0 40 0 550,clip,width=\hsize]{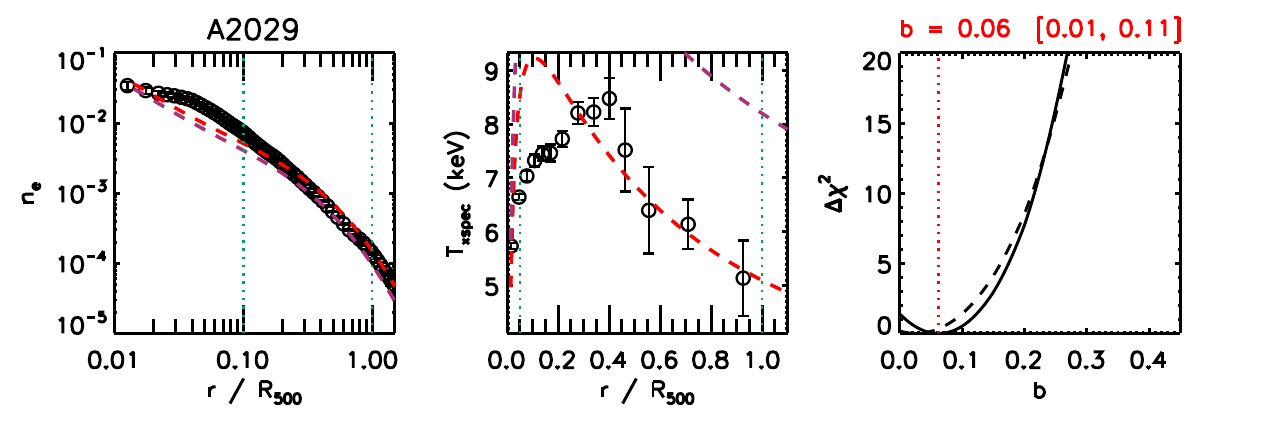} 
\includegraphics[page=8,trim=0 40 0 550,clip,width=\hsize]{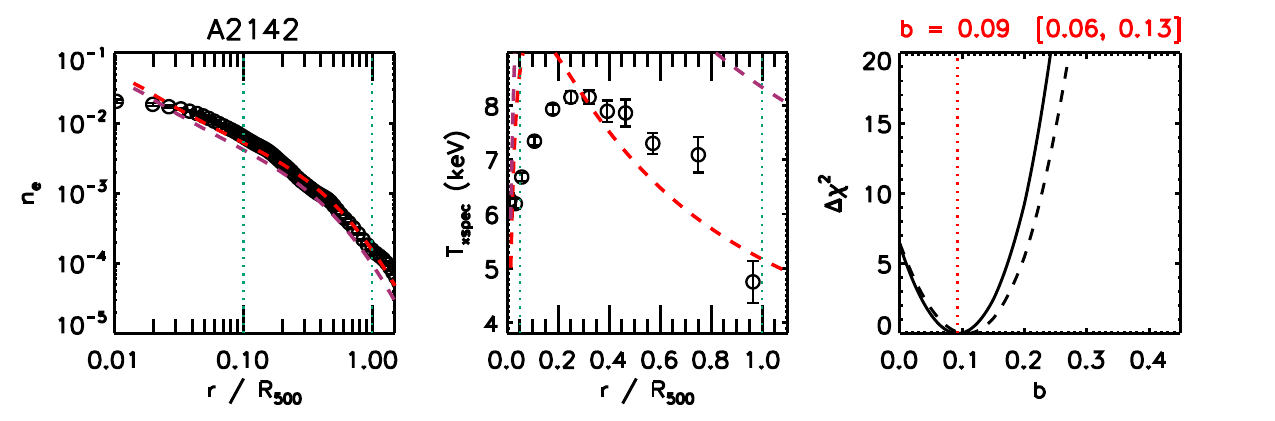} 
\caption{Radial profiles of the gas density (left) and spectroscopic temperature (center) recovered for the four X-COP objects and, overplotted, the predictions obtained by assuming $b=0$ (red dashed line) and $b=0.6$ (purple dashed line). (Right panel) $\Delta \chi^2$ for the analysis using only the $T$ profile (dashed line) and jointly with the $n$ profile (solid line).
} \label{fig:xcop}
\end{figure*}

\begin{figure*}[ht]
\centering
\includegraphics[page=10,trim=0 40 0 550,clip,width=\hsize]{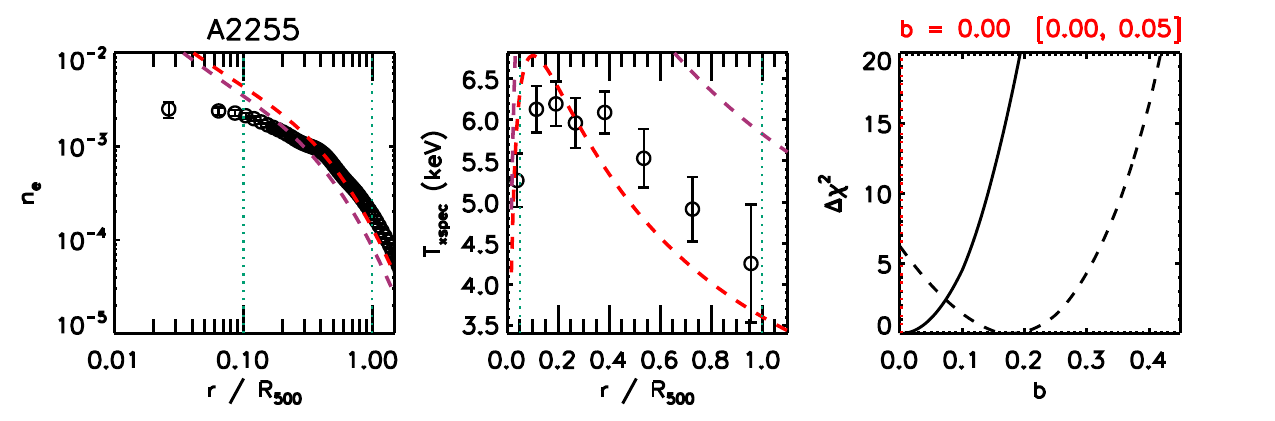} 
\includegraphics[page=12,trim=0 40 0 550,clip,width=\hsize]{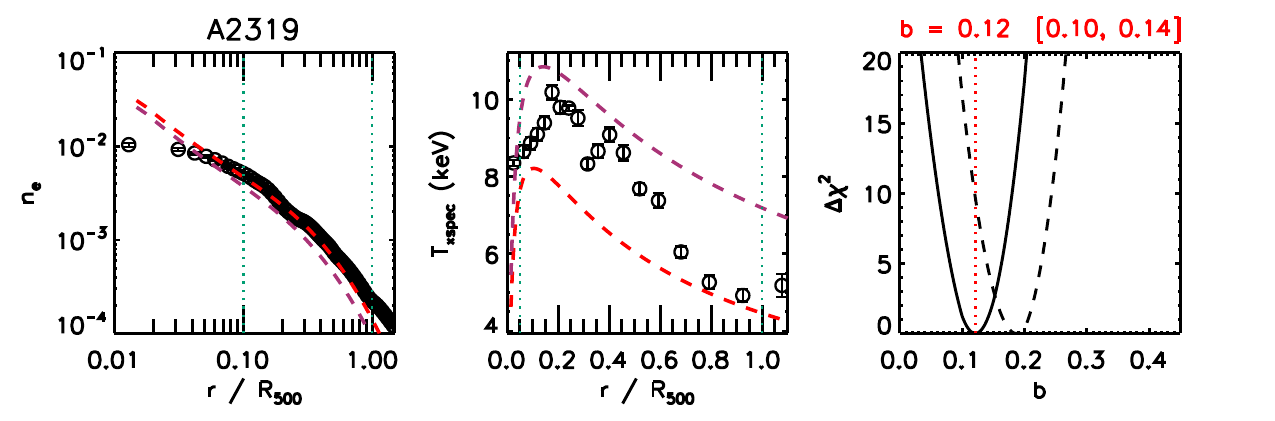} 
\includegraphics[page=14,trim=0 40 0 550,clip,width=\hsize]{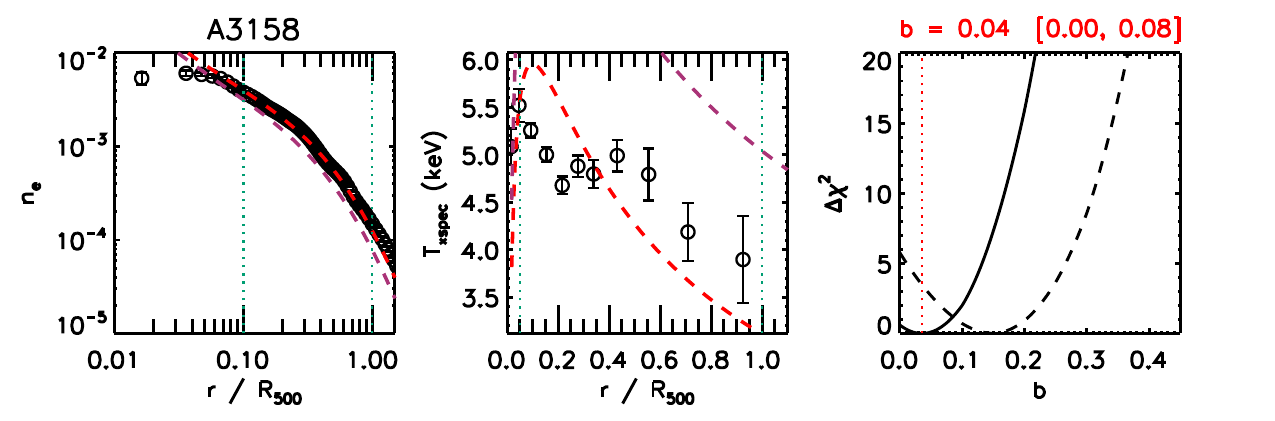} 
\includegraphics[page=16,trim=0 40 0 550,clip,width=\hsize]{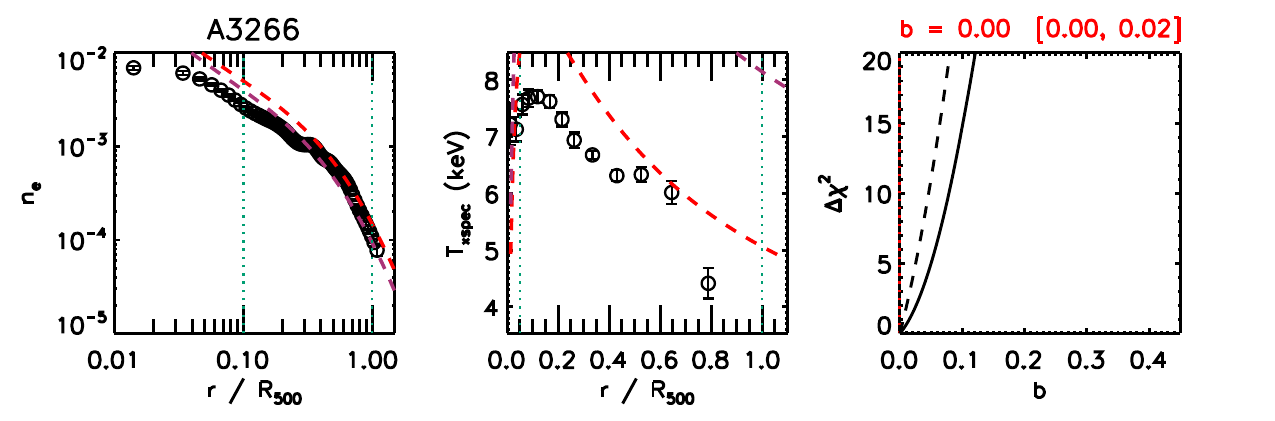} 
\caption{Continued.
} 
\end{figure*}

\begin{figure*}[ht]
\centering
\includegraphics[page=18,trim=0 40 0 550,clip,width=\hsize]{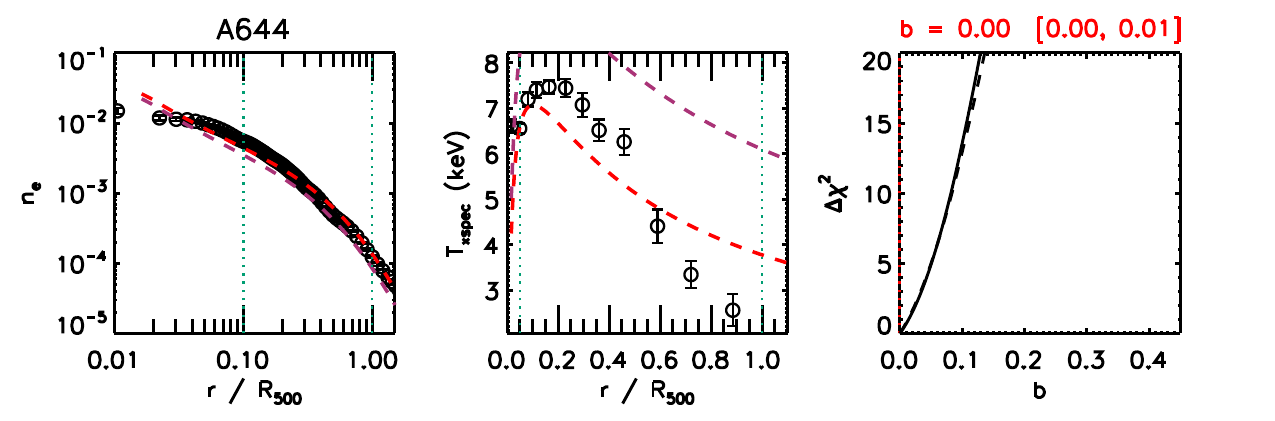} 
\includegraphics[page=20,trim=0 40 0 550,clip,width=\hsize]{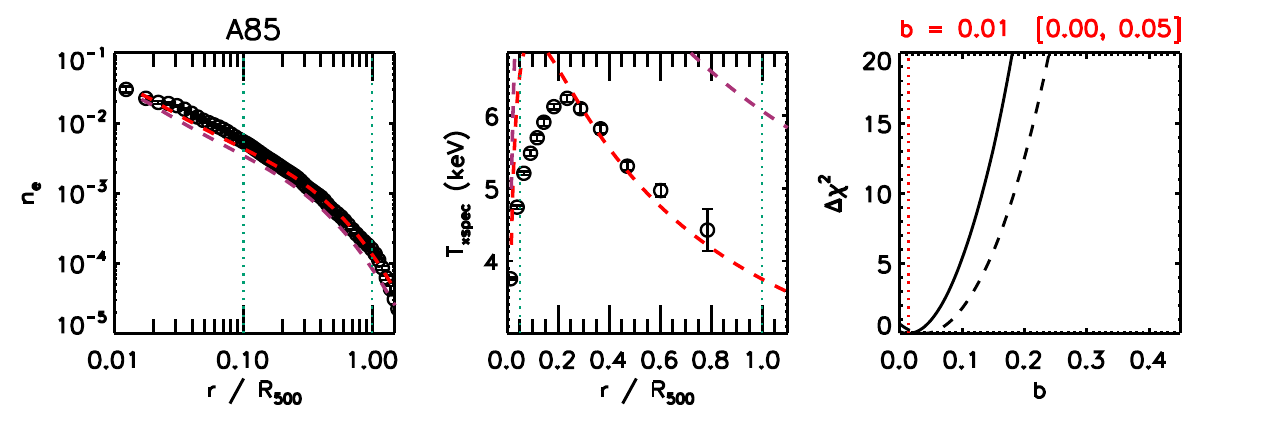} 
\includegraphics[page=22,trim=0 40 0 550,clip,width=\hsize]{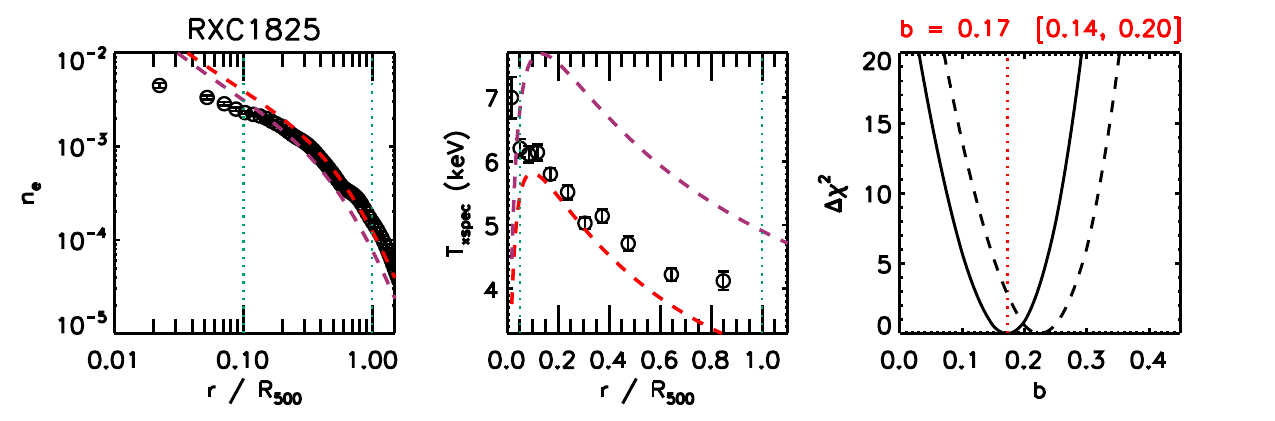} 
\includegraphics[page=24,trim=0 40 0 550,clip,width=\hsize]{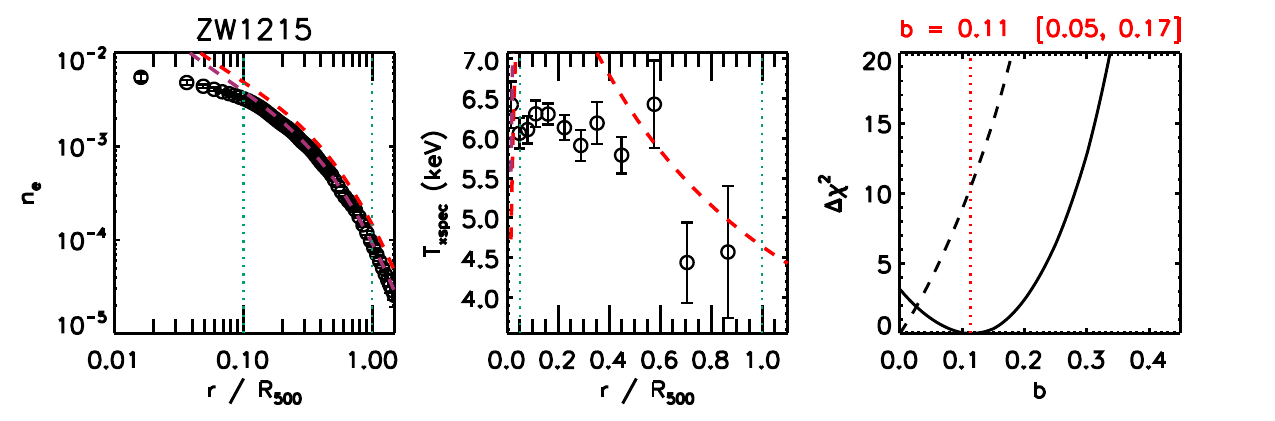} 
\caption{Continued.
} 
\end{figure*}

\end{appendix}
\end{document}